\documentclass[twocolumn, showpacs,amsmath,amssymb]{revtex4-1}
\usepackage{graphicx}% Include figure files
\usepackage{dcolumn}% Align table columns on decimal point
\usepackage{bm}% bold math
\usepackage{stackrel}% http://ctan.org/pkg/stackrel
\usepackage{amsmath}
\usepackage{amsfonts}
\usepackage{amssymb}
\usepackage{amsthm}
\usepackage{epstopdf}
\usepackage{longtable}
\usepackage{array}
\usepackage{pifont}
\usepackage{latexsym}
\usepackage[latin1]{inputenc}
\usepackage{bm}
\usepackage{textcomp}
\usepackage{subfigure}
\usepackage{braket}
\usepackage{wasysym}
\usepackage{ulem}
\usepackage{color, soul}
\usepackage{pifont}
\usepackage{multirow}
 \pdfoutput=1

\newcolumntype{.}{D{.}{.}{-1}}

\begin{document}

\title{Universal ultra-robust interrogation protocol with zero probe-field-induced frequency shift for quantum clocks and high-accuracy spectroscopy}

\author{T. Zanon-Willette$^{1}$\footnote{E-mail address: thomas.zanon@upmc.fr}, R. Lefevre$^{1}$}
\affiliation{$^{1}$ LERMA, Observatoire de Paris, PSL Research University, CNRS, Sorbonne Universités, UPMC Univ. Paris 06, F-75005, Paris, France}
\author{A.V. Taichenachev$^{2,3}$, V.I. Yudin$^{2,3}$}
\affiliation{$^{2}$ Novosibirsk State University, ul. Pirogova 2, Novosibirsk 630090, Russia}
\affiliation{$^{3}$ Institute of Laser Physics, SB RAS, pr. Akademika Lavrent'eva 13/3, Novosibirsk 630090, Russia}
\date{\today}
\date{\today}

\preprint{APS/123-QED}

\begin{abstract}
Optical clock interrogation protocols, based on laser-pulse spectroscopy, are suffering from probe-induced frequency shifts and their variations induced by laser power.
Original Hyper-Ramsey probing scheme, which was proposed to alleviate those issues, does not fully eliminate the shift, especially when decoherence and relaxation by spontaneous emission or collisions are present.
We propose to solve the fundamental problem of frequency shifts induced by laser probe by deriving the exact canonical form of a multi-pulse generalized Hyper-Ramsey (GHR) resonance, including decoherence and relaxation. We present a universal interrogation protocol based on composite laser-pulses spectroscopy with phase-modulation
eliminating probe-induced frequency shifts at all orders in presence of various dissipative processes.
Unlike frequency shifts extrapolation-based methods, a universal interrogation protocol based on $\pm\pi/4$ and $\pm3\pi/4$ phase-modulated resonances is proposed which does not compromise the stability of the optical clock while maintaining an ultra-robust error signal gradient in presence of substantial uncompensated ac Stark-shifts. Such a scheme can be implemented in two flavours: either by inverting clock state initialization or by pulse order reversal even without a perfect quantum state initialization.
This universal interrogation protocol can be applied to atomic, molecular and nuclear frequency metrology, mass spectrometry and to the field of precision spectroscopy. It might be designed using magic-wave induced transitions, two-photon excitation and magnetically-induced spectroscopy or it might even be implemented with quantum logic gate circuit and qubit entanglement.
\end{abstract}


\maketitle

\section{INTRODUCTION}

\indent Atomic optical clocks are recognized to be ideal platforms for highly accurate frequency measurements, leading to very stringent tests
of fundamental physical theories \cite{Ludlow:2015}, such as relativity \cite{Chou:2010-1,Dzuba:2017}, detection of gravitational waves \cite{Kolkowitz:2016}, possible variation of fundamental constants with time \cite{Uzan:2015}, or search for dark matter \cite{Derevianko:2015}.
Depending on the selected atomic species used to achieve stable and accurate optical frequency standards, single trapped ion clocks \cite{Chou:2010-2,Margolis:2009} and neutral atoms lattice clocks \cite{Ye:2008,Derevianko:2011,Katori:2011} have been characterized over many years, reducing systematic uncertainties to a fractional frequency change well below $10^{-16}$, surpassing current microwave atomic frequency standards.
These promising standards are based on ultra-narrow electric-dipole-forbidden transitions. For ions, examples are spin-forbidden transitions using quantum logic spectroscopy \cite{Schmidt:2005}, or electric-quadrupole or octupole transitions, as in the single $^{171}$Yb$^{+}$ ion clock which has recently demonstrated a relative $3\times10^{-18}$ systematic uncertainty \cite{Huntemann:2016}. Optical lattice clocks with alkaline-earth-like atoms are based on a doubly forbidden transition weakly allowed in fermions (odd isotopes) by a level mixing due to the hyperfine structure.
 $^{171}$Yb and $^{87}$Sr optical lattice clocks are now reaching relative stabilities in the $10^{-16}$ range \cite{Schioppo:2017} and relative accuracies of $2\times10^{-18}$ \cite{Nicholson:2015}, potentially leading to a redefinition of the second for the next decade \cite{LeTargat:2013,Riehle:2015}.
Strongly forbidden transitions, with vanishing spin-orbit coupling due to zero nuclear spin, have been studied more recently in bosonic species ($^{88}$Sr, $^{174}$Yb, $^{24}$Mg), but they require a two-photon excitation technique \cite{Santra:2005,Zanon-Willette:2006} or a magnetically-induced spectroscopy \cite{Taichenachev:2006,Barber:2006,Baillard:2007,Kulosa:2015}, which are both limited by important AC Stark-shifts or Zeeman frequency shifts. Because the quest for extreme precision in ultra-high resolution spectroscopy is still progressing, it will ultimately require new laser stabilization protocols, reducing systematic uncertainties to very low levels, pushing precision even further. Among these uncertainties, frequency shifts from the laser-probe itself are always present and might become a severe limitation for the next generation of fermionic and bosonic quantum clocks with fractional frequency change below $10^{-18}$.

Ramsey spectroscopy \cite{Ramsey:1956} has been first modified by including a frequency step during the laser pulses in order to compensate the probe-induced frequency shift \cite{Yudin:2009}. However, when the shift is not fully compensated, a frequency shift remains, with a linear dependence to the error on the compensation.
Then, composite laser pulses techniques so-called Hyper-Ramsey (HR) spectroscopy, previously developed in nuclear magnetic resonance and quantum computation \cite{Levitt:1986,Vandersypen:2005,Braun:2014}, were applied with electromagnetic phase-modulated resonances \cite{Ramsey:1951,Morinaga:1989,Letchumanan:2006} in order to provide non-linear elimination of residual uncompensated light-shift contributions and laser power variations \cite{Yudin:2010,Zanon:2015,Zanon-Willette:2016-1,Zanon-Willette:2016-2}.
Such a HR spectroscopy has been successfully applied on the ultra-narrow electric octupole transition of the single $^{171}$Yb$^{+}$ ion, reducing ac Stark-shifts by four orders of magnitude and was proven to be shielded from small pulse area variations \cite{Huntemann:2012-2}.
To completely remove the third-order weak dependence of the HR clock frequency shift on light-shift uncompensated parts, a modified Hyper-Ramsey technique (MHR) was experimentally implemented within a bosonic $^{88}$Sr lattice clock, demonstrating the suppression of the $2\times10^{-13}$ probe Stark shifts to below $10^{-16}$, drastically expanding the acceptance bandwidth of imperfect shift compensation \cite{Hobson:2016}.
\begin{figure}[t!!]
\center
\resizebox{9.5cm}{!}{\includegraphics[angle=0]{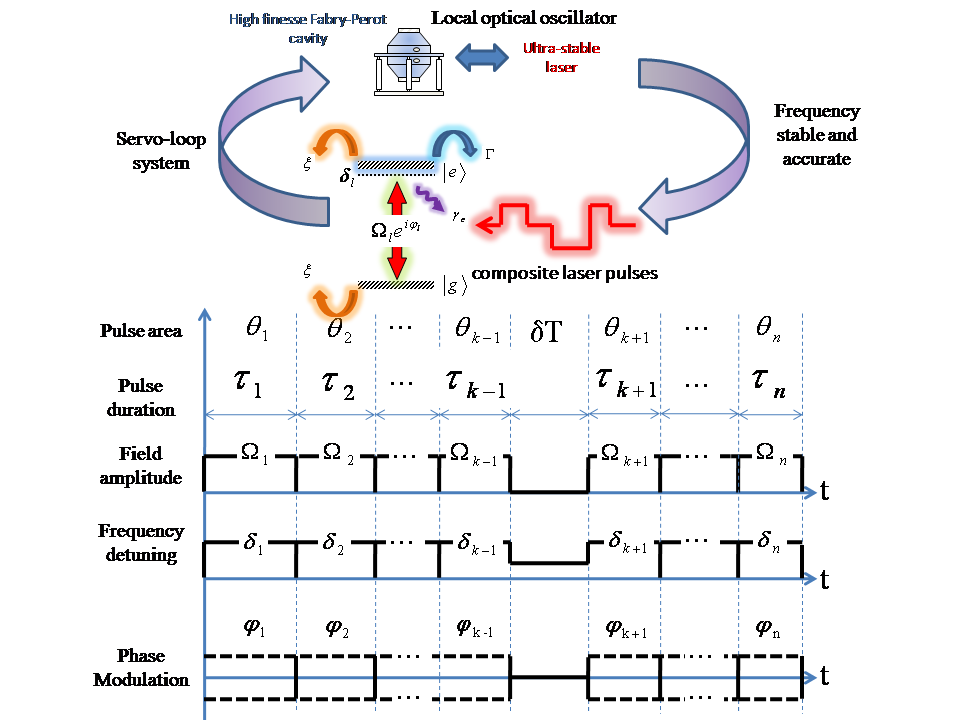}}
\caption{(color online) Composite laser pulse spectroscopy probing a fermionic or a bosonic clock transition perturbed by dissipative processes. Optical pulses are defined by a generalized area $\theta_{l}$ ($l=\textup{1,2..,k,..,n}$), the frequency detuning $\delta_{l}$, the field amplitude $\Omega_{l}e^{i\varphi_{l}}$ including a phase-step modulation $\varphi_{l}$, a pulse duration $\tau_{l}$ and a single free evolution time T applied somewhere at the desired $l=\textup{k}$ pulse. The general clock frequency detuning is defined by $\delta_{l}=\delta-\Delta_{l}$ where a residual error in pre-compensation of laser probe induced frequency shift is $\Delta_{l}$. The laser induced decoherence is called $\gamma_{c}$, relaxation by spontaneous emission is labeled $\Gamma$ and $\xi$ is the relaxation rate of the population difference due to collisions.}
\label{fig:composite-pulses}
\end{figure}
However, it has been pointed out that the reliability of interrogation schemes against uncompensated probe
frequency shifts and laser power variations might be severely limited by decoherence, compromising the improvements of further metrological performances \cite{Yudin:2016}.

We manage to overcome this fundamental obstacle by building an ultra-robust clock-laser stabilization scheme taking into account atomic decoherence and
relaxation by both spontaneous emission and weak collisions. The error signal is synthesized by repeating and combining several atomic population
excitation fraction measurements, interleaved by a controllable population inversion between clock states.
The paper is organized as follows: We begin by presenting the two-level optical Bloch-equations which are used to describe coherent interaction between laser and atoms
including several dissipative processes which may disrupt the clock transition.
The Bloch vector resulting of a multi-pulse generalized hyper-Ramsey (GHR) resonance is first expressed in a canonical form with a clock frequency shift. Note that such our formal analytic (GHR) resonance pattern can integrate additional NMR rotation composite pulse protocols \cite{Levitt:1986} to remove any potential additional errors if desired.
The corresponding error signal line-shape is then derived and the associated clock frequency shift is obtained by a combination of phase-modulated (GHR) resonances.
We introduce a general 2D diagram approach for frequency shift reconstruction allowing a global map analysis of decoherence and relaxation effects.
The main part of this paper is dedicated to a universal laser interrogation protocol using a combination of multiple (GHR) error signals based on $\pi/4,3\pi/4$ phase-steps
and quantum state initialization generating a laser frequency locking-point which is immune to probe-induced frequency shifts.
We finally explore and compare the sensitivity of the original HR interrogation protocol, such as applied to the single ion $^{171}$Yb$^+$ clock \cite{Huntemann:2012-2},
to our universal laser frequency stabilization technique for different radiative configurations of a two-level clock transition.
\begin{table*}[t!!]
\centering%
\caption{Composite laser-pulses interrogation protocols ignoring dissipative processes. The clock frequency shift including
residual error in pre-compensation of probe-induced frequency shifts $\Delta$ is given by $\delta\widetilde{\nu}(\Delta/\Omega)$. Pulse area $\boldsymbol{\theta_{l}}$ is given in degrees and phase-steps $\varphi_{l+},\varphi_{l-}$ are indicated in subscript-brackets with radian unit. The standard Rabi frequency for all pulses is $\Omega=\pi/2\tau$ where $\tau$ is the pulse duration reference. Free evolution appears at index $\textup{k}=2$, denoted $\theta_{\textup{k}}=\delta\textup{T}$. Reverse composite pulses protocols are denoted by $(\dagger)$.}
\label{protocol-table-1}
\renewcommand{\arraystretch}{1.7}
\begin{tabular}{|c|c|c|}
\hline
\hline
protocols [\textup{refs}] & composite pulses $\boldsymbol{\theta_{l}}$$_{(\varphi_{l+},\varphi_{l-})}$  & $\delta\widetilde{\nu}(\Delta/\Omega)$  \\
\hline
     R \cite{Ramsey:1956} & \begin{tabular}{c}
$\boldsymbol{90}_{(\frac{\pi}{2},-\frac{\pi}{2})}\dashv\delta\textup{T}\vdash\boldsymbol{90}_{(0,0)}$ \\
$(\dagger)$ $\boldsymbol{90}_{(0,0)}\dashv\delta\textup{T}\vdash\boldsymbol{90}_{(-\frac{\pi}{2},\frac{\pi}{2})}$
                                \end{tabular}
     &  $\frac{1}{\pi\textup{T}}\frac{\Delta}{\Omega}$ \\

\hline
\hline

HR \cite{Yudin:2010,Zanon:2015}   & \begin{tabular}{c}
$\boldsymbol{90}_{(\frac{\pi}{2},-\frac{\pi}{2})}\dashv\delta\textup{T}\vdash\boldsymbol{180}_{(\pi,\pi)}\boldsymbol{90}_{(0,0)}$ \\
$(\dagger)$  $\boldsymbol{90}_{(0,0)}\boldsymbol{180}_{(\pi,\pi)}\dashv\delta\textup{T}\vdash\boldsymbol{90}_{(-\frac{\pi}{2},\frac{\pi}{2})}$
                                   \end{tabular}
 &  $\frac{4}{\pi\textup{T}}\left(\frac{\Delta}{\Omega}\right)^{3}$  \\

\hline
\hline

     MHR \cite{Hobson:2016} & \begin{tabular}{c}
$\boldsymbol{90}_{(\frac{\pi}{2},0)}\dashv\delta\textup{T}\vdash\boldsymbol{180}_{(\pi,\pi)}\boldsymbol{90}_{(0,-\frac{\pi}{2})}$ \\
$(\dagger)$  $\boldsymbol{90}_{(-\frac{\pi}{2},0)}\boldsymbol{180}_{(\pi,\pi)}\dashv\delta\textup{T}\vdash\boldsymbol{90}_{(0,\frac{\pi}{2})}$
      \end{tabular}
   &  0   \\

\hline
\hline

   GHR$(\frac{\pi}{4})$ \cite{Zanon-Willette:2016-1} & \begin{tabular}{c}
$\boldsymbol{90}_{(0,0)}\dashv\delta\textup{T}\vdash\boldsymbol{180}_{(\frac{\pi}{4},-\frac{\pi}{4})}\boldsymbol{90}_{(0,0)}$  \\
$(\dagger)$  $\boldsymbol{90}_{(0,0)}\boldsymbol{180}_{(-\frac{\pi}{4},\frac{\pi}{4})}\dashv\delta\textup{T}\vdash\boldsymbol{90}_{(0,0)}$
         \end{tabular}
   &  0  \\

\hline
\hline

    GHR$(\frac{3\pi}{4})$ \cite{Zanon-Willette:2016-1} & \begin{tabular}{c}
$\boldsymbol{90}_{(0,0)}\dashv\delta\textup{T}\vdash\boldsymbol{180}_{(3\frac{\pi}{4},-3\frac{\pi}{4})}\boldsymbol{90}_{(0,0)}$ \\
$(\dagger)$  $\boldsymbol{90}_{(0,0)}\boldsymbol{180}_{(-\frac{3\pi}{4},\frac{3\pi}{4})}\dashv\delta\textup{T}\vdash\boldsymbol{90}_{(0,0)}$
            \end{tabular}
    &  0 \\
\hline
\hline
\end{tabular}
\end{table*}

\section{CANONICAL FORM FOR ANALYTICAL MULTI-PULSE (GHR) RESONANCE EXPRESSION}

\indent To design a universal interrogation protocol for fermions and bosons, we first derive the exact analytical expression of a phase-modulated generalized Hyper-Ramsey (GHR) resonance along with the clock frequency shift expression, including dissipative processes \cite{Tabatchikova:2013,Tabatchikova:2015}. The atomic transition, shown in Fig.~\ref{fig:composite-pulses}, includes a decoherence term $\gamma_{c}$, a spontaneous emission rate denoted
$\Gamma$ and an excited state population relaxation $\xi$ induced by weak collisions. Bloch variables are used to describe the fraction of population excitation after
successive optical composite pulses with area $\theta_{l}$ indexed by $l=\textup{1,...,k,...,n}$, including a free evolution time T at index $l=\textup{k}$.
Light pulse duration $\tau_{l\neq k}$, Rabi frequency $\Omega_{l}$, laser detuning $\delta_{l}$, and phase $\varphi_{l}$ of the coherent electromagnetic field can be modified independently over the entire sequence.
The general set of time-dependent optical Bloch equations for a two-level $\{|\textup{g}\rangle,|\textup{e}\rangle\}$ quantum system for the $l$-th pulse is given by
\cite{Torrey:1949,Jaynes:1955,Allen:1975,Schoemaker:1978,Cohen-Tannoudji:1992,Berman:2011}:
\begin{equation}
\left\lbrace
\begin{split}
\dot{\textup{U}}_{l}=&-\gamma_{c}~\textup{U}_{l} + \delta_{l}~\textup{V}_{l} - \Omega_{l}\sin\varphi_{l}~\textup{W}_{l},\\
\dot{\textup{V}}_{l}=&-\delta_{l}~\textup{U}_{l} -\gamma_{c}~\textup{V}_{l} + \Omega_{l}\cos\varphi_{l}~\textup{W}_{l},\\
\dot{\textup{W}}_{l}=&~\Omega_{l}\sin\varphi_{l}~\textup{U}_{l} - \Omega_{l}\cos\varphi_{l}~\textup{V}_{l}-(\Gamma+2\xi)~\textup{W}_{l}-\Gamma.\\
\end{split}\right.
\label{set-Bloch-DM}
\end{equation}
where $\delta_{l}=\delta-\Delta_{l}$ is the generalized clock frequency detuning, with $\delta$ being the laser frequency detuning from the unperturbed clock resonance.
A frequency offset is added to the detuning $\delta$ during all light pulses, but not during the free evolution time $T$, to bring back the observed central fringe near $\delta=0$ \cite{Yudin:2009}. $\Delta_{l}$ is the part of the frequency shift non compensated by the applied frequency offset.
Optical coherence and population difference are related to density matrix elements by $\textup{U}_{l}\equiv\rho_{\textup{ge}}+\rho_{\textup{ge}}^{*}$, $\textup{V}_{l}\equiv i(\rho_{\textup{ge}}-\rho_{\textup{ge}}^{*})$ and $\textup{W}_{l}\equiv\rho_{\textup{ee}}-\rho_{\textup{gg}}$. Population conservation is given by the relation $\rho_{\textup{gg}}+\rho_{\textup{ee}}=1$.
The complete three-vector components $\textup{M}(\theta_{l})\equiv( \textup{U}(\theta_{l}),\textup{V}(\theta_{l}),\textup{W}(\theta_{l}))$ solution to the previous set of equations is \cite{Jaynes:1955,Schoemaker:1978}:
\begin{equation}
\begin{split}
\textup{M}(\theta_{l})
=\textup{R}(\theta_{l})\left[\textup{M}_{l}(0)-\textup{M}_{l}(\infty)\right]+\textup{M}_{l}(\infty).
\end{split}
\label{Jaynes-solution}
\end{equation}
where we introduce for convenience a generalized pulse area $\theta_{l}=\omega_{l}\tau_{l}$ and a generalized Rabi frequency $\omega_{l}$  (see appendix A for all definitions).
The rotation matrix $\textup{R}(\theta_{l})$, taking into account decoherence and relaxation terms, is written as follows:
\begin{equation}
\begin{split}
\textup{R}(\theta_{l})&=e^{-\gamma_{c}\tau_{l}}e^{-\beta_{l}\tau_{l}},\\
\beta_{l}&=\left(
                                                                                          \begin{array}{ccc}
                                                                                            0 & \delta & -\Omega_{l}\sin\varphi_{l} \\
																				            -\delta & 0  & \Omega_{l}\cos\varphi_{l}\\
																			              	\Omega_{l}\sin\varphi_{l} & -\Omega_{l}\cos\varphi_{l} & \Delta\gamma
                                                                                          \end{array}\right),
\end{split}
\label{rotation-matrix}
\end{equation}
with $\Delta\gamma=\gamma_{c}-(\Gamma+2\xi)$. $\textup{M}_{l}(0)\equiv(\textup{U}_{l}(0),\textup{V}_{l}(0),\textup{W}_{l}(0))$ stands for the system's state before the $l$-th pulse.
The exponential matrix $\textup{R}(\theta_{l})$ can be exactly expressed as a square  matrix of time-dependent matrix elements $\textup{R}_{\textup{mn}}(\theta_{l})$
($\textup{m,n}=1,2,3$) (refer to the appendix A for all details).
Steady-state solutions $\textup{M}_{l}(\infty)\equiv(\textup{U}_{l}(\infty),\textup{V}_{l}(\infty),\textup{W}_{l}(\infty))$ are directly obtained by switching off time-dependent derivatives in Eq.~(\ref{set-Bloch-DM}) for the three vector-components.
The free evolution matrix $\textup{R}(\theta_{\textup{k}})$ at index $l=\textup{k}$  without laser field reduces to:
\begin{equation}
\begin{split}
\textup{R}(\theta_{\textup{k}}=\delta\textup{T})=e^{-\gamma_{c}\textup{T}}\left(
\begin{array}{ccc}
\cos\delta\textup{T} & \sin\delta\textup{T} & 0 \\
-\sin\delta\textup{T} & \cos\delta\textup{T} & 0\\
0 & 0 & e^{\Delta\gamma\textup{T}}
\end{array}\right).
\end{split}
\label{free-rotation-matrix}
\end{equation}
The corresponding stationary solution $\textup{M}_{\textup{k}}(\infty)\equiv(\textup{U}_{\textup{k}}(\infty), \textup{V}_{\textup{k}}(\infty),\textup{W}_{\textup{k}}(\infty))$ is also found by switching off the laser field $\Omega_{\textup{k}}=0$ in Eq.~(\ref{set-Bloch-DM}) during free evolution time.

The complete solution of Bloch-vector components for a full sequence can ultimately be expressed in a reduced canonical form:
\begin{equation}
\textup{M}(\theta_{\textup{1}},...,\theta_{\textup{n}})\equiv\textup{A}+\textup{B}(\Phi)\cos(\delta\textup{T}+\Phi),
\label{eq:canonical-form}
\end{equation}
which is the generalization to $n$ pulses of the expression established for $n=3$ \cite{Zanon:2015}.
The offset term $\textup{A}$ is given by:
\begin{equation}
\begin{split}
\textup{A}=&\sum_{\textup{p=k+1}}^{\textup{n}}\left[ \left(\overleftarrow{\prod_{l=\textup{p}}^{\textup{n}}}\textup{R}(\theta_{l})\right)\left(\textup{M}_{\textup{p}-1}(\infty)-\textup{M}_{\textup{p}}(\infty)\right)\right]\\
&+\textup{M}_{\textup{n}}(\infty)+\textup{h}~e^{-(\Gamma+2\xi)\textup{T}}\begin{pmatrix}
\tilde{\textup{R}}_{13}\\
\tilde{\textup{R}}_{23}\\
\tilde{\textup{R}}_{33}  \end{pmatrix}.
\end{split}
\label{reduced-components}
\end{equation}
The amplitude term $\textup{B}(\Phi)$ components are given by:
\begin{equation}
\textup{B}_{i}(\Phi_{i})=e^{-\gamma_{c}\textup{T}}|\textup{C}_{i}|\sqrt{1+\tan^{2}\Phi_{i}}\hspace{0.25cm} i\in\{1,2,3\},
\end{equation}
and the phase-shift term $\Phi$ components are written as:
\begin{equation}
\begin{split}
\Phi_{i}&=-\arctan\left[\textup{S}_{i}/\textup{C}_{i}\right]\hspace{0.25cm} i\in\{1,2,3\},\\
\textup{S}&\equiv\begin{pmatrix}
\tilde{\textup{R}}_{11}~\textup{g}-\tilde{\textup{R}}_{12}~\textup{f} \\
\tilde{\textup{R}}_{21}~\textup{g}-\tilde{\textup{R}}_{22}~\textup{f} \\
\tilde{\textup{R}}_{31}~\textup{g}-\tilde{\textup{R}}_{32}~\textup{f}
 \end{pmatrix},\hspace{0.5cm}
\textup{C}\equiv\begin{pmatrix}
\tilde{\textup{R}}_{11}~\textup{f}+\tilde{\textup{R}}_{12}~\textup{g} \\
\tilde{\textup{R}}_{21}~\textup{f}+\tilde{\textup{R}}_{22}~\textup{g} \\
\tilde{\textup{R}}_{31}~\textup{f}+\tilde{\textup{R}}_{32}~\textup{g}
\end{pmatrix},
\end{split}
\label{GHR-phase-shift}
\end{equation}
where $\tilde{\textup{R}}_{\textup{mn}}$ ($\textup{m,n=1,2,3}$) are the matrix elements of the compiled matrix $\tilde{\textup{R}}$ and ($\textup{f,g,h}$) components given by:
\begin{equation}
\begin{split}
\tilde{\textup{R}}&=\overleftarrow{\prod_{l=\textup{k}+1}^{\textup{n}}}\textup{R}(\theta_{l}),\\
\begin{pmatrix}
\textup{f} \\ \textup{g} \\ \textup{h}
\end{pmatrix}&=\sum_{\textup{p}=1}^{\textup{k}}\left(\overleftarrow{\prod_{l=\textup{p}}^{\textup{k}-1}}\textup{R}(\theta_{l})\right)(\textup{M}_{\textup{p}-1}(\infty)-\textup{M}_{\textup{p}}(\infty)).
\end{split}
\label{}
\end{equation}
where backward arrows indicate a matrix product from right to left with growing indices.
The generalized Hyper-Ramsey canonical expression describing the population transfer from $|\textup{g}\rangle$ to $|\textup{e}\rangle$ clock states is
given by the third component of the Bloch variables. Various composite pulses protocols and their time-reversed counterparts reported in Table.~\ref{protocol-table-1} can be simulated using Eq.~(\ref{eq:canonical-form}).

A high-order expression of the clock frequency shift $\delta\nu\approx-\Phi|_{\delta\rightarrow0}/2\pi\textup{T}$
affecting the extremum of the central fringe pattern of the GHR resonance from Eq.~(\ref{eq:canonical-form}) is presented in appendix B.
Ignoring dissipative processes, analytical expressions have already been derived for Ramsey and Hyper-Ramsey protocols in Refs \cite{Zanon:2015,Zanon-Willette:2016-2}.
The typical non-linear response of a GHR resonance line-shape to probe-induced frequency shifts is usually asymmetric leading to
off-center line locking when a laser frequency modulation technique is applied.
In the next section, we present the phase-step modulation of the resonance shape which eliminates
the effect of that asymmetry on the true position of the central fringe.

\section{ERROR SIGNAL GENERATION WITH PHASE-MODULATED (GHR) RESONANCE}

\indent A laser frequency stabilization scheme based on anti-symmetric laser phase-steps is able to synthesize a dispersive error signal
locking the laser frequency to the center of the perturbed clock transition \cite{Ramsey:1951,Letchumanan:2006}. This technique is applied by measuring experimentally the population transfer $P_{|\textup{g}\rangle\mapsto|\textup{e}\rangle}$ between clock states. The phase-modulated Ramsey scheme requires
the relative phase of the second optical Ramsey pulse to be shifted by $±\pi/2$ with respect to the first pulse.

For simplicity, we now focus on the third Bloch variable component related to population difference and we will omit indices in subsequent expressions. 
The error signal $\Delta\textup{E}$ for a particular protocol is built by taking the difference between two Bloch-vector components $\textup{M}(\theta_{1},...,\theta_{\textup{n}})$ with
appropriate phase-steps modulation $_{(\varphi_{l+},\varphi_{l-})}$ of a specified pulse area $\theta_{l}$. The resulting line-shape for population transfer between clock states is:
\begin{equation}
\begin{split}
\Delta\textup{E}&\equiv\textup{M}(\theta_{1},...,\theta_{\textup{n}})(\varphi_{l+})-\textup{M}(\theta_{1},...,\theta_{\textup{n}})(\varphi_{l-}),\\
&=\left(\textup{P}_{|\textup{g}\rangle\mapsto|\textup{e}\rangle}(\varphi_{l+})-\textup{P}_{|\textup{g}\rangle\mapsto|\textup{e}\rangle}(\varphi_{l-})\right).
\end{split}
\end{equation}
The new phase-modulated lineshape can also be rewritten in yet another phasor canonical form as:
\begin{equation}
\begin{split}
\Delta\textup{E}\equiv\tilde{\textup{A}}+\tilde{\textup{B}}(\tilde{\Phi})\cos(\delta\textup{T}+\tilde{\Phi}).
\end{split}
\label{eq:error-signal}
\end{equation}
where offset $\tilde{\textup{A}}$, amplitude $\tilde{\textup{B}}$ and phase-shift $\tilde{\Phi}$ are explicitly given in appendix C.
The error signal shape for the third component related to the population difference exhibits a dispersive feature versus clock frequency detuning, unlike the GHR resonance curve \cite{Zanon:2015}.
From the condition $\Delta\textup{E}|_{\delta=\delta\widetilde{\nu}}=0$ due to imperfect probe-induced frequency shift compensation, it is straightforward to
derive a new analytical form of the frequency-shifted locking-point $\delta\widetilde{\nu}$ as:
\begin{equation}
\begin{split}
\delta\widetilde{\nu}=\frac{1}{2\pi\textup{T}}\left(-\tilde{\Phi}|_{\delta\rightarrow0}\pm\arccos\left[-\frac{\tilde{\textup{A}}|_{\delta\rightarrow0}}{\tilde{\textup{B}}(\tilde{\Phi})|_{\delta\rightarrow0}}\right]\right).
\end{split}
\label{eq:clock-frequency-shift}
\end{equation}
The robustness of various error signals to a modification of pulse area and uncompensated frequency shifts has already been numerically studied in detail when decoherence is non negligible \cite{Yudin:2016}. The fundamental consequence for all optical interrogation schemes is a rapid loss of the laser frequency locking-point stability inducing, for example, additional constraints concerning the MHR protocol \cite{Yudin:2016}.
\begin{figure*}[t!!]
\center
\resizebox{6.5cm}{!}{\includegraphics[angle=0]{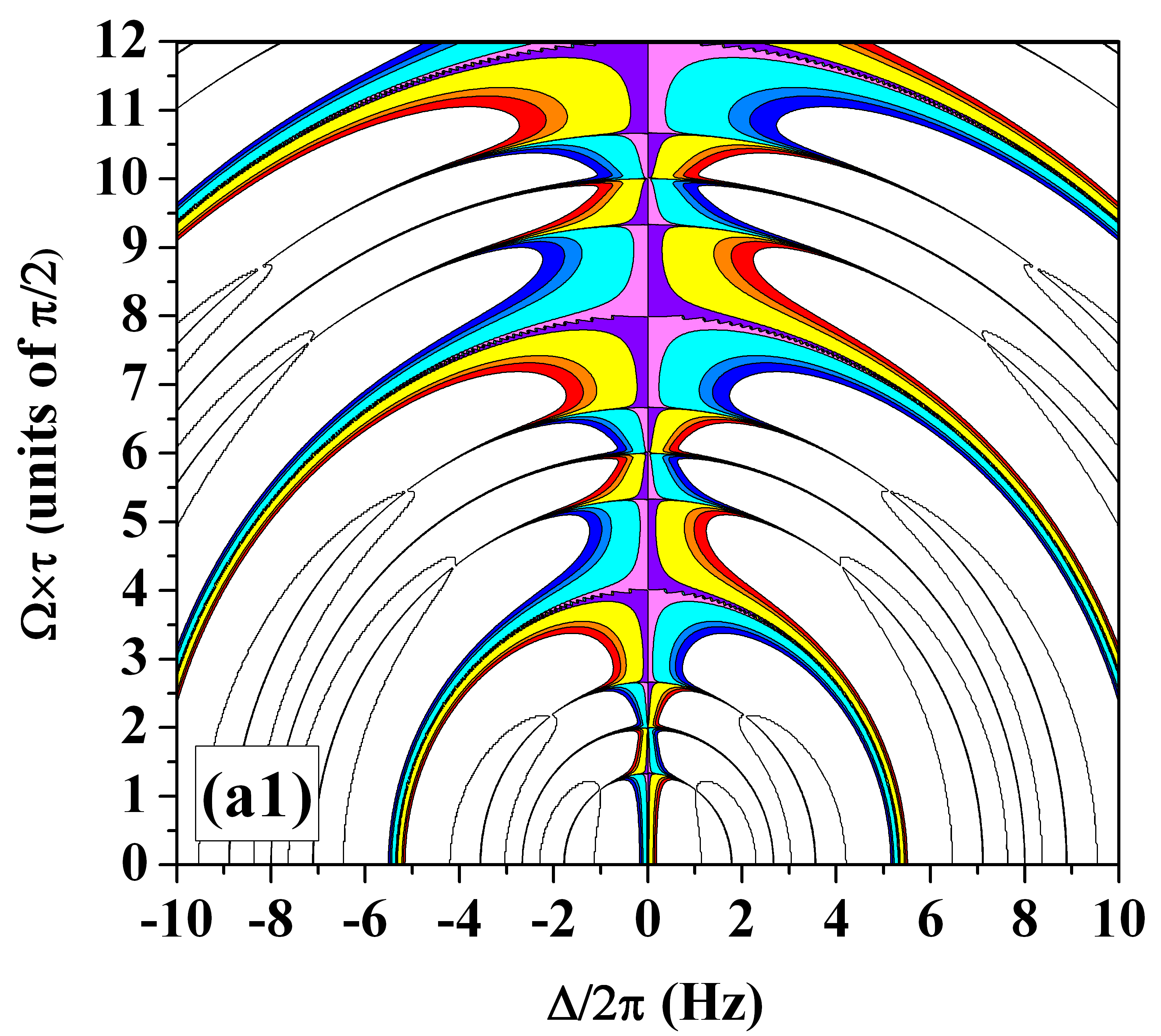}}\resizebox{6.5cm}{!}{\includegraphics[angle=0]{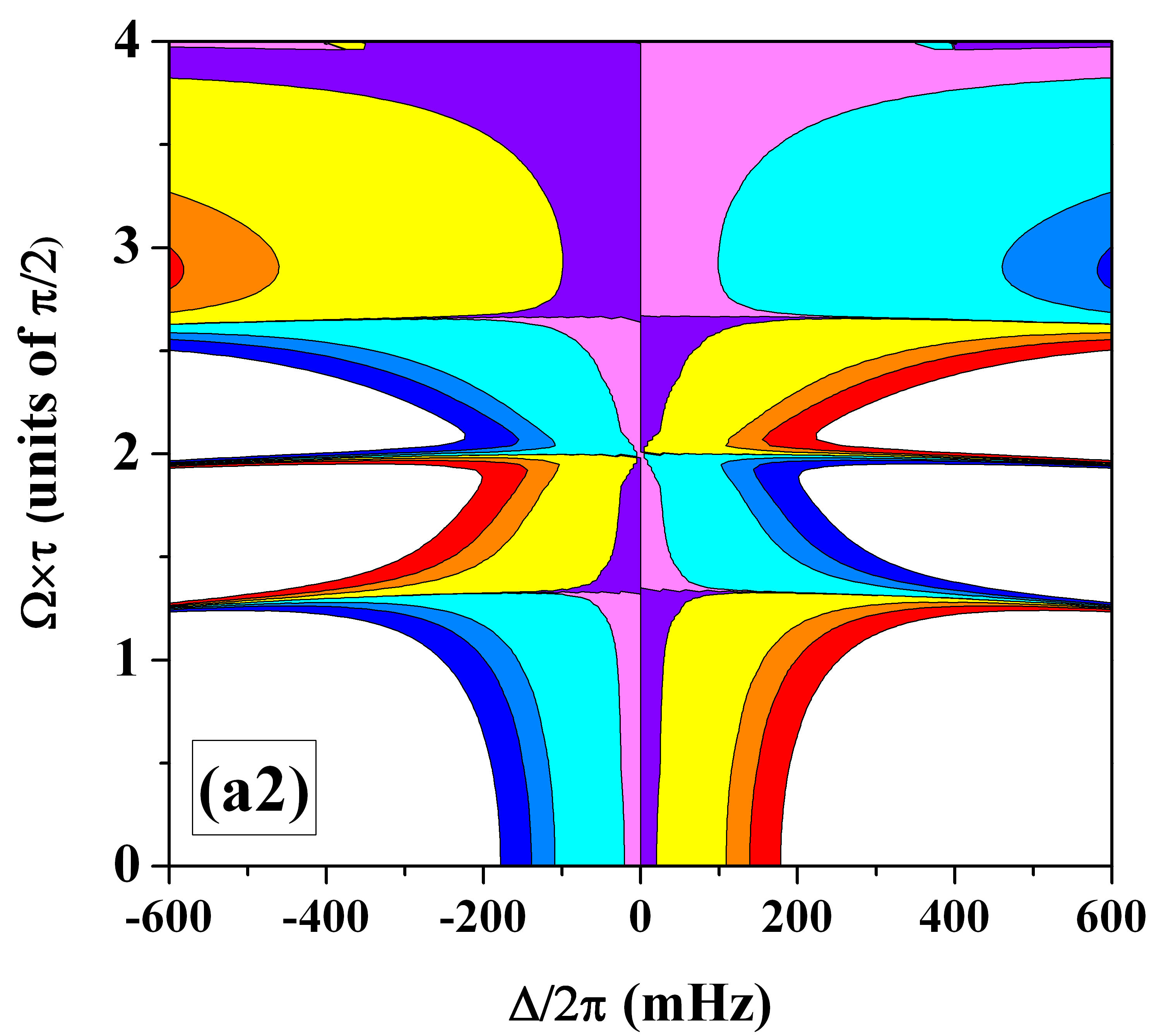}}
\resizebox{6.5cm}{!}{\includegraphics[angle=0]{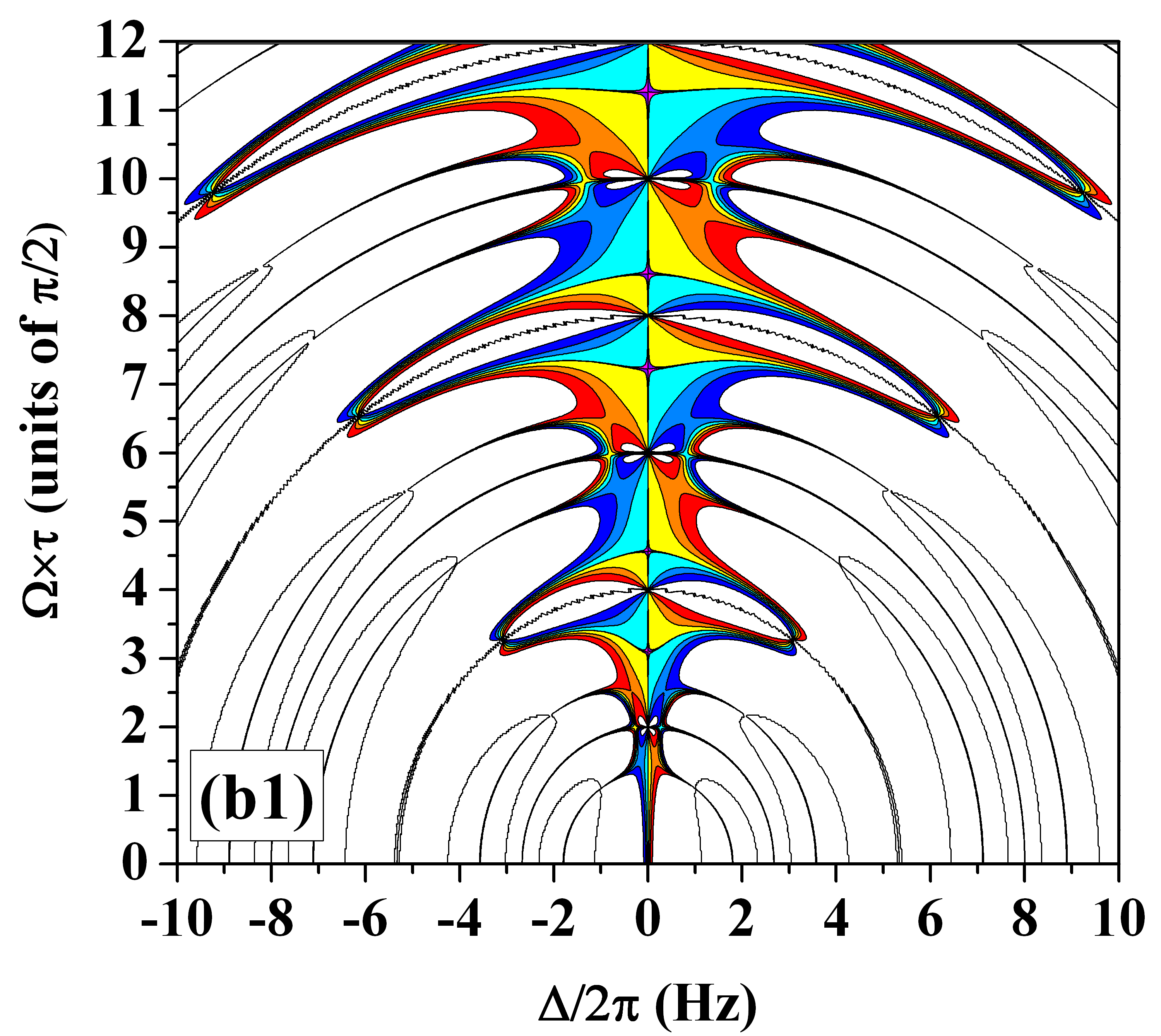}}\resizebox{6.5cm}{!}{\includegraphics[angle=0]{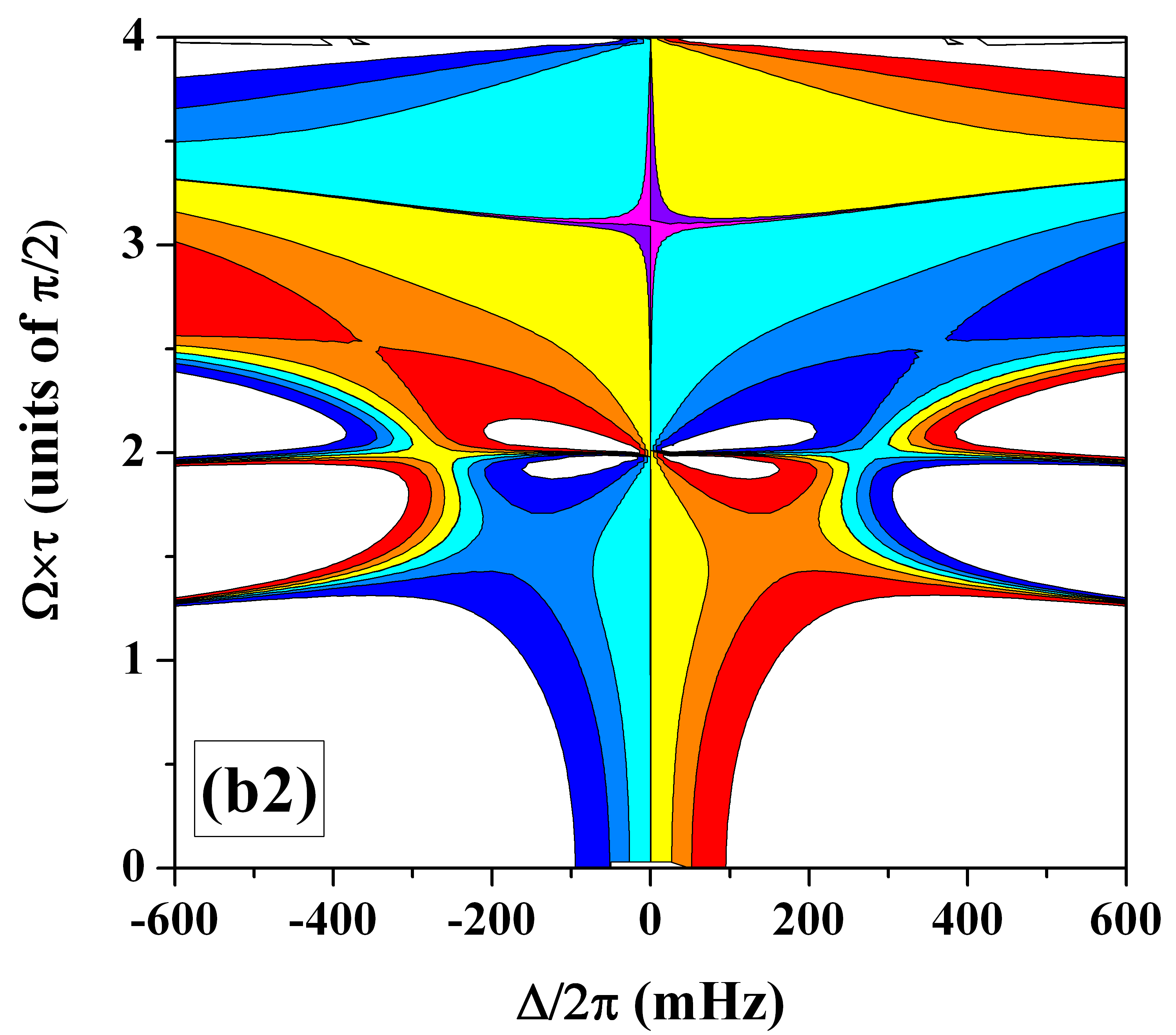}}
\resizebox{6.5cm}{!}{\includegraphics[angle=0]{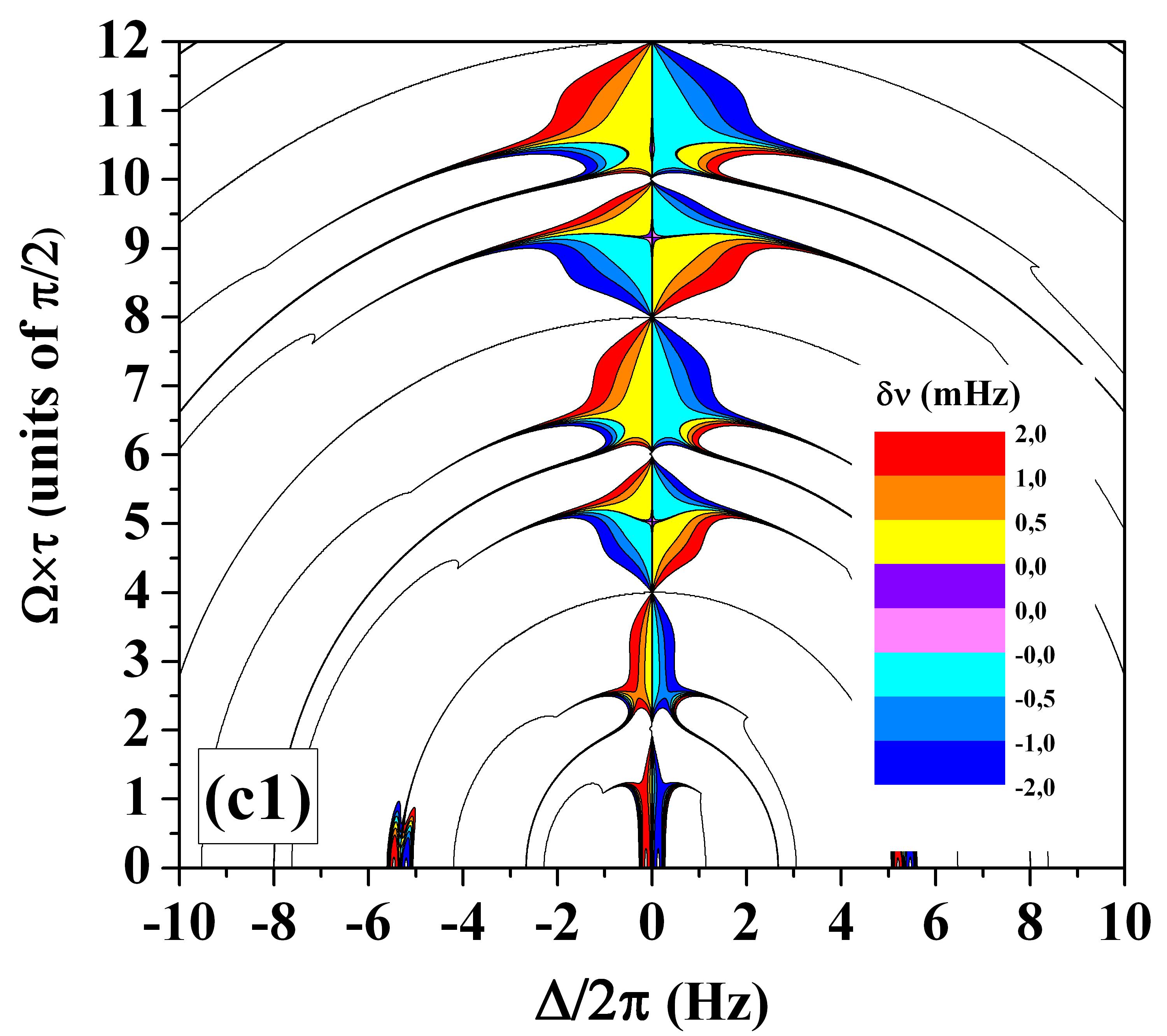}}\resizebox{6.5cm}{!}{\includegraphics[angle=0]{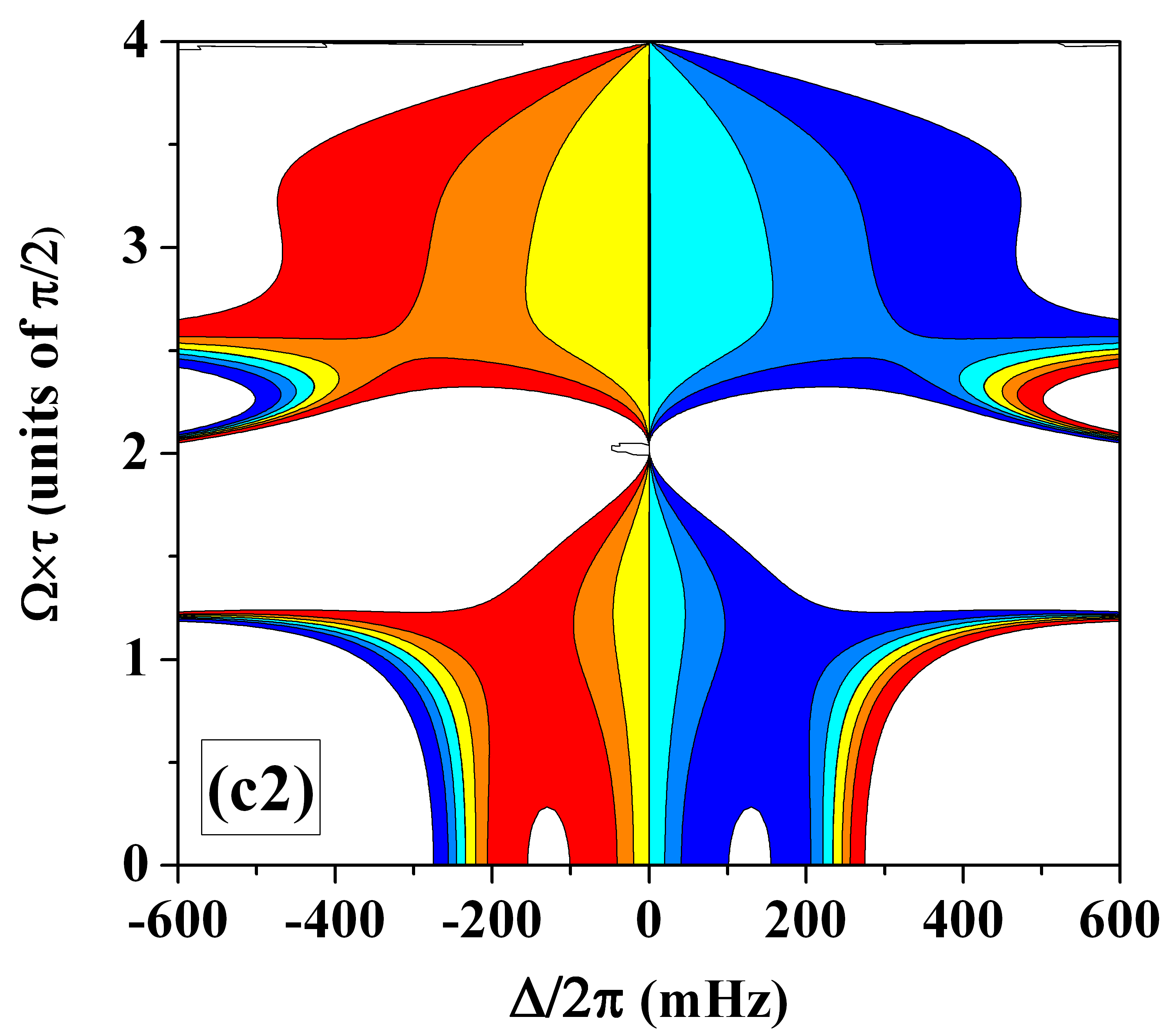}}
\caption{(color online) 2D contour and density plot diagrams of the $\delta\widetilde{\nu}[\textup{HR}]$ clock frequency shift based on Eq.~(\ref{eq:clock-frequency-shift}) versus
uncompensated frequency shifts $\Delta/2\pi$ (horizontal axis) and pulse area variation $\Omega\tau$ (vertical axis). Left
graphs are over a large detuning acceptance bandwidth and right graphs are expanded between $\pi/2$ and $3\pi/2$ pulse areas.
(a1,a2) Ideal case. (b1,b2) Decoherence $\gamma_{c}=2\pi\times50$~mHz. (c1,c2) Decoherence and relaxation $\gamma_{c}=2\pi\times50$~mHz,
$\Gamma=2\pi\times100$~mHz. The standard Rabi frequency for all pulses is $\Omega=\pi/2\tau$ where $\tau$ is the pulse duration reference. Pulse duration reference is set to $\tau=3/16$~s, free evolution time is $\textup{T}=2$~s and uncompensated frequency-shift is $\Delta_{l}\equiv\Delta$ ($l=1,3,4$).}
\label{fig:2D-map-HR-diagrams}
\end{figure*}
To explore in more depth the instability of frequency locking-points caused by dissipative processes, clock frequency shifts for various interrogation protocols have been extracted from general offset and amplitude terms established in the previous section, Eq.~(\ref{eq:error-signal}) and Eq.~(\ref{eq:clock-frequency-shift}).
They are investigated with the help of 2D contour and density plot diagrams presented in the next section.

\section{2D DIAGRAMS FOR CLOCK FREQUENCY SHIFT RECONSTRUCTION}

\begin{figure*}[t!!]
\center
\resizebox{6.5cm}{!}{\includegraphics[angle=0]{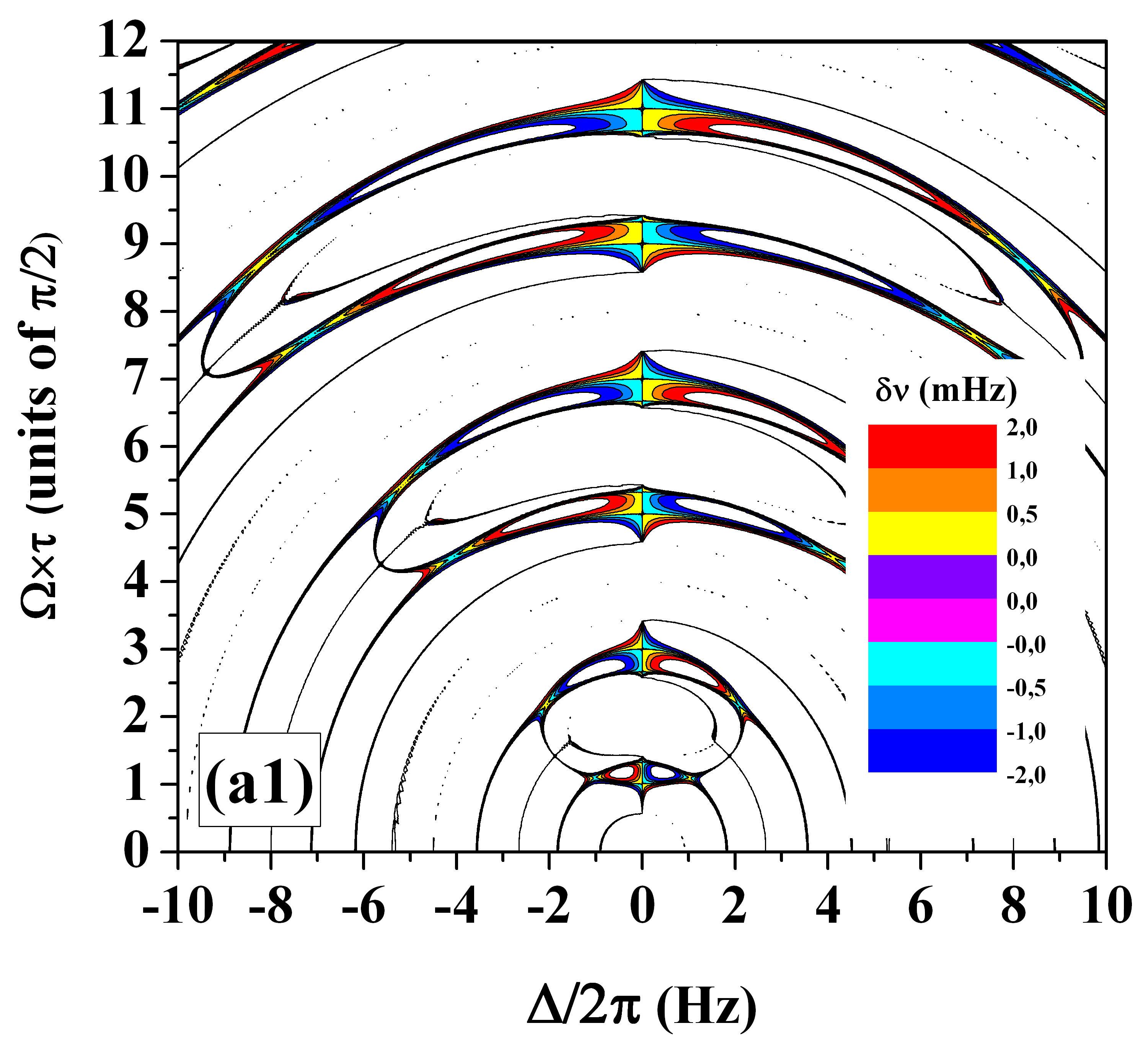}}\resizebox{6.5cm}{!}{\includegraphics[angle=0]{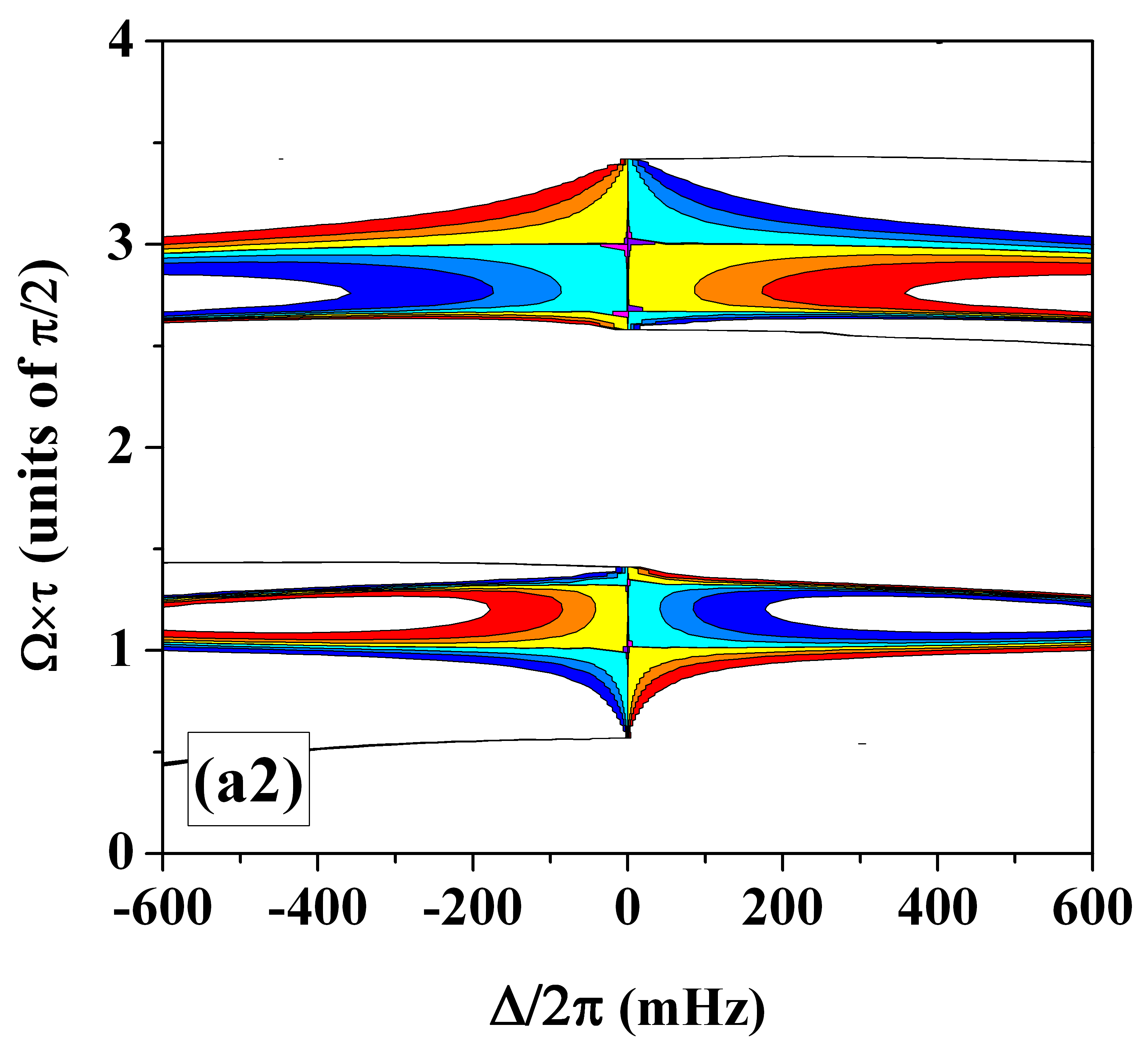}}
\resizebox{6.5cm}{!}{\includegraphics[angle=0]{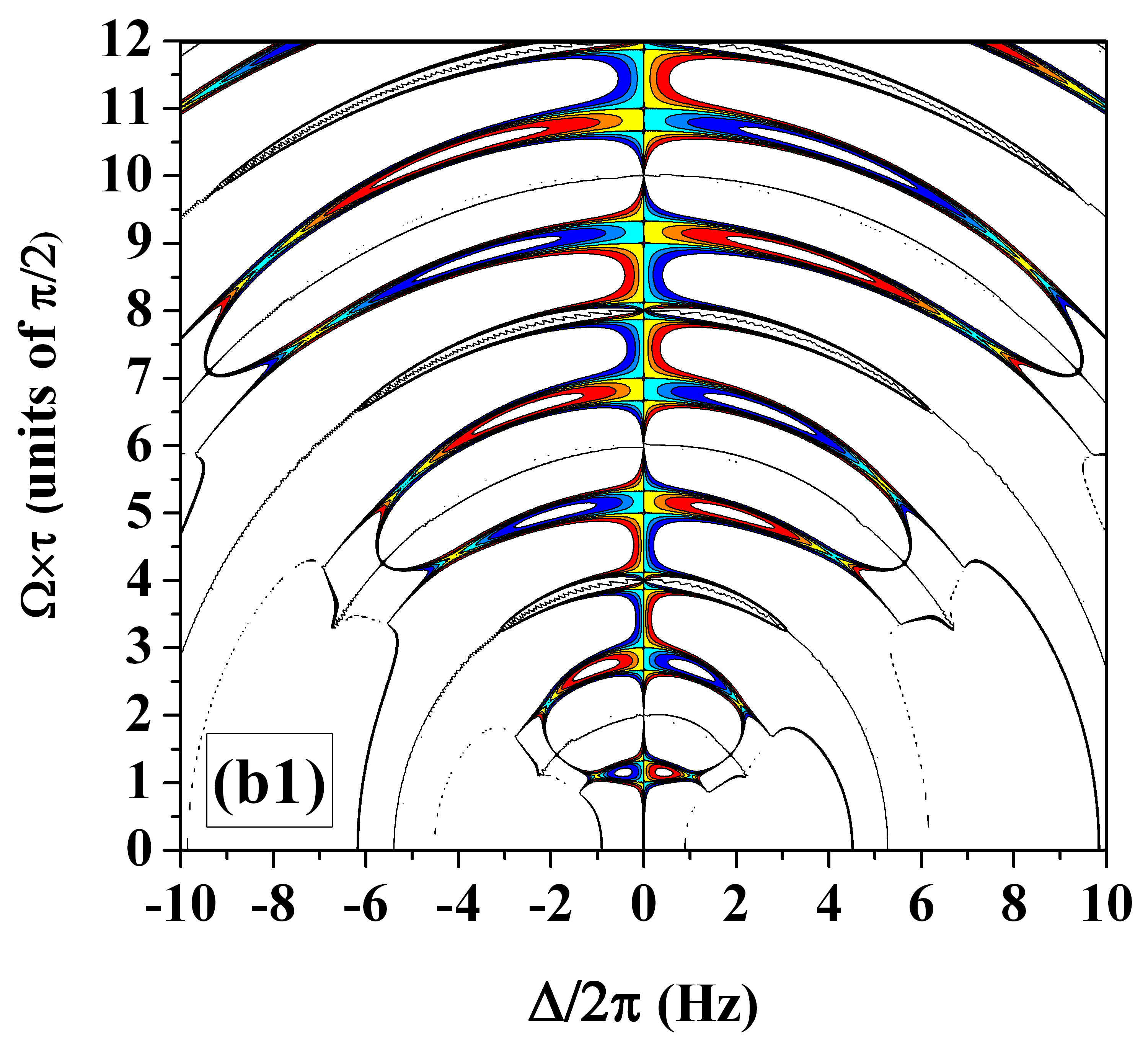}}\resizebox{6.5cm}{!}{\includegraphics[angle=0]{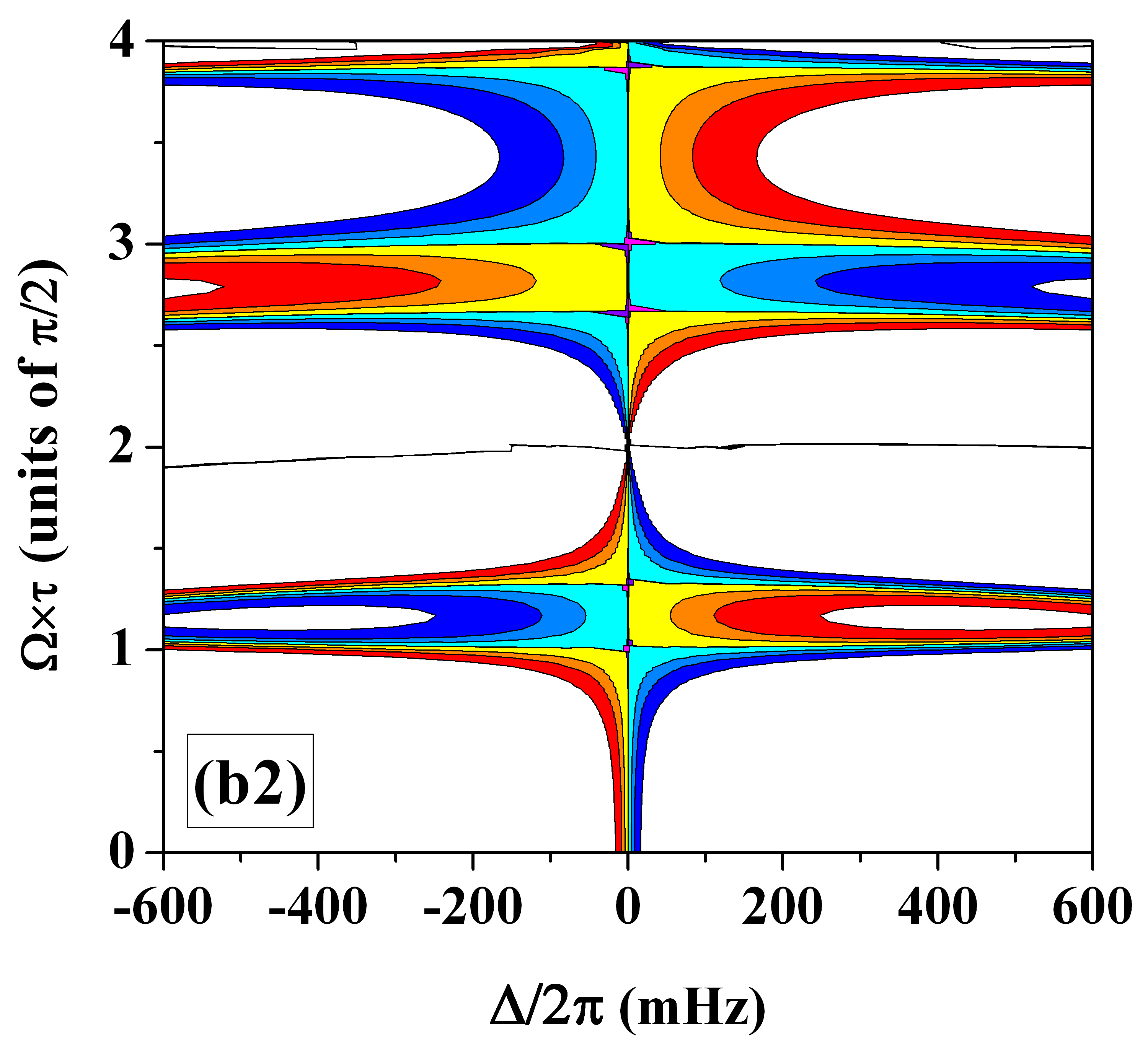}}
\resizebox{6.5cm}{!}{\includegraphics[angle=0]{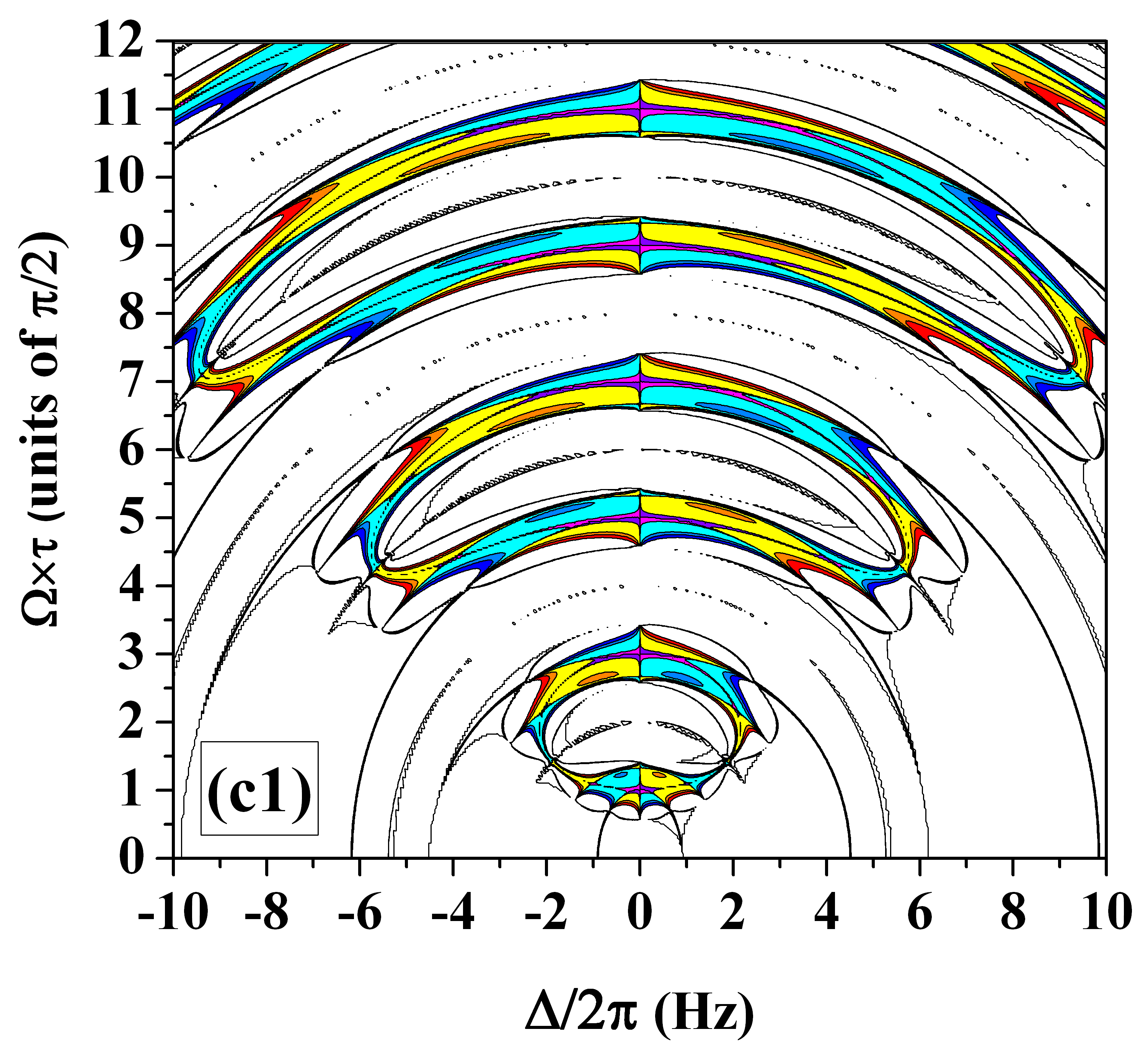}}\resizebox{6.5cm}{!}{\includegraphics[angle=0]{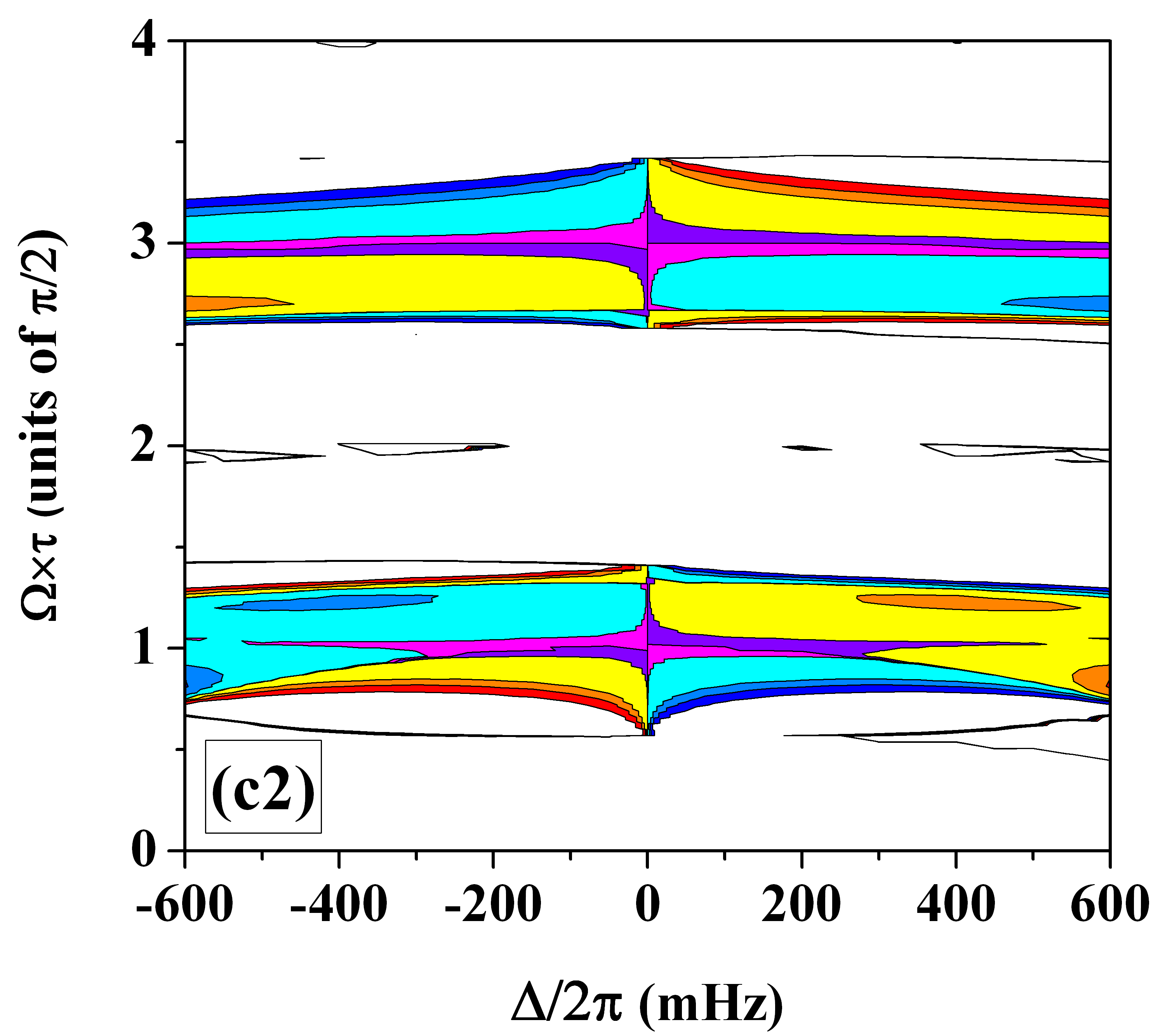}}
\caption{(color online) 2D contour and density plot diagrams of the $\delta\widetilde{\nu}[\textup{GHR}(\pi/4)]$ and $\delta\widetilde{\nu}[\textup{GHR}(3\pi/4)]$ clock frequency shifts based on Eq.~(\ref{eq:clock-frequency-shift}) under decoherence $\gamma_{c}=2\pi\times50$~mHz versus uncompensated frequency shifts $\Delta/2\pi$ (horizontal axis) and pulse area variation $\Omega\tau$ (vertical axis). Left graphs are over a large detuning acceptance bandwidth and right graphs are expanded between $\pi/2$ and $3\pi/2$ pulse areas.
(a1,a2) $\delta\widetilde{\nu}[\textup{GHR}(\pi/4)]$ diagram. (b1,b2) $\delta\widetilde{\nu}[\textup{GHR}(3\pi/4)]$ diagram. (c1,c2) Synthetic frequency-shift
$\delta\widetilde{\nu}[\textup{syn}]=\frac{1}{2}(\delta\widetilde{\nu}[\textup{GHR}(\pi/4)]+\delta\widetilde{\nu}[\textup{GHR}(3\pi/4)])$. Other parameters are identical to Fig.~\ref{fig:2D-map-HR-diagrams}.}
\label{fig:2D-map-GHR-diagrams}
\end{figure*}

\indent The influence of decoherence or relaxation by spontaneous emission on HR and GHR probing schemes is analyzed using
2D contour and density plot diagrams shown in Figs.~\ref{fig:2D-map-HR-diagrams} and \ref{fig:2D-map-GHR-diagrams}.
All clock-frequency shifts $\delta\widetilde{\nu}$ are plotted using Eq.~(\ref{eq:clock-frequency-shift})
versus uncompensated frequency shifts and large pulse area variations. Because ac Stark-shifts increasing quadratically with pulse area might still be manageable by applying a larger laser frequency-step for pre-compensation of the central fringe frequency-shift \cite{Yudin:2009}, diagrams are also exploring regions of several $\pi/2$ laser pulse area units.
Note that the error signal contrast is always maximized for odd values of multiples of $\Omega\tau=\pi/2$ pulses and vanishing for even values. Colored values of clock-frequency shifts have been deliberately limited between -2~mHz and +2~mHz for constrain below $10^{-18}$ relative accuracy. The white background represents some regions where the residual shift is exceeding a few $10^{-18}$ levels of relative accuracy.

The dependence of the HR error signal $\Delta\textup{E}[\textup{HR}]$ on uncontrollable modifications of laser parameters, ignoring dissipative processes,
is presented in Fig.~\ref{fig:2D-map-HR-diagrams}(a1) and (a2). The 2D contour and density plots exhibit some stable regions where the third-order
dependence of the clock-shift $\delta\widetilde{\nu}[\textup{HR}]$ is well below 500~$\mu$Hz over 100~mHz of uncompensated
frequency shifts (pink and violet region along the vertical axis). The clock-frequency-shift compensation can be made more robust over a wider range
of residual frequency shifts by increasing the pulse area from $\pi/2$ to a magic value near $2.95\pi/2$ as shown in Fig.~\ref{fig:2D-map-HR-diagrams}(a2).
At this particular value, all contour plots (black thin isoclinic lines delimiting regions) present vanishing first-order derivative versus pulse area variation
making the frequency locking-point even more stable to small laser power modification. Noteworthy frequency locking-points are also observed near the value
of $1.2\pi/2$ or near $2.6\pi/2$. Around these values, the clock frequency shift is changing abruptly from positive to negative values for small errors in
compensation of probe-induced frequency shifts.
When there is decoherence, a modification of the $\delta\widetilde{\nu}[\textup{HR}]$ clock-frequency shift is observed in Fig.~\ref{fig:2D-map-HR-diagrams}(b1,b2) leading to a linear
increase of the shift up to 2~mHz over 100~mHz of uncompensated frequency shifts. However, a small frequency stability island (small pink and violet region) emerges
in Fig.~\ref{fig:2D-map-HR-diagrams}(b2) for a pulse area near $\sim3.25\pi/2$.
When decoherence and relaxation by spontaneous emission are both present as shown in Fig.~\ref{fig:2D-map-HR-diagrams}(c1,c2), the clock frequency shift
is reversed with a negative slope of 2~mHz over 400~mHz of uncompensated frequency shifts.\\
\begin{figure*}[t!!]
\begin{minipage}[c]{0.45\linewidth}
\centering
\resizebox{9.0cm}{!}{\includegraphics[angle=0]{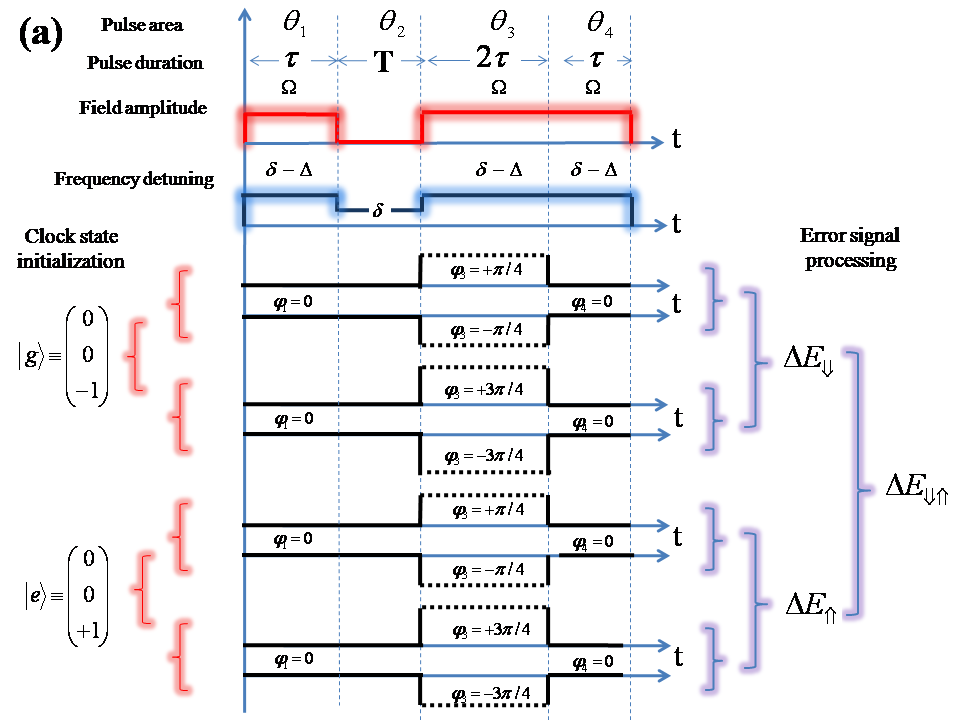}}
\resizebox{9.0cm}{!}{\includegraphics[angle=0]{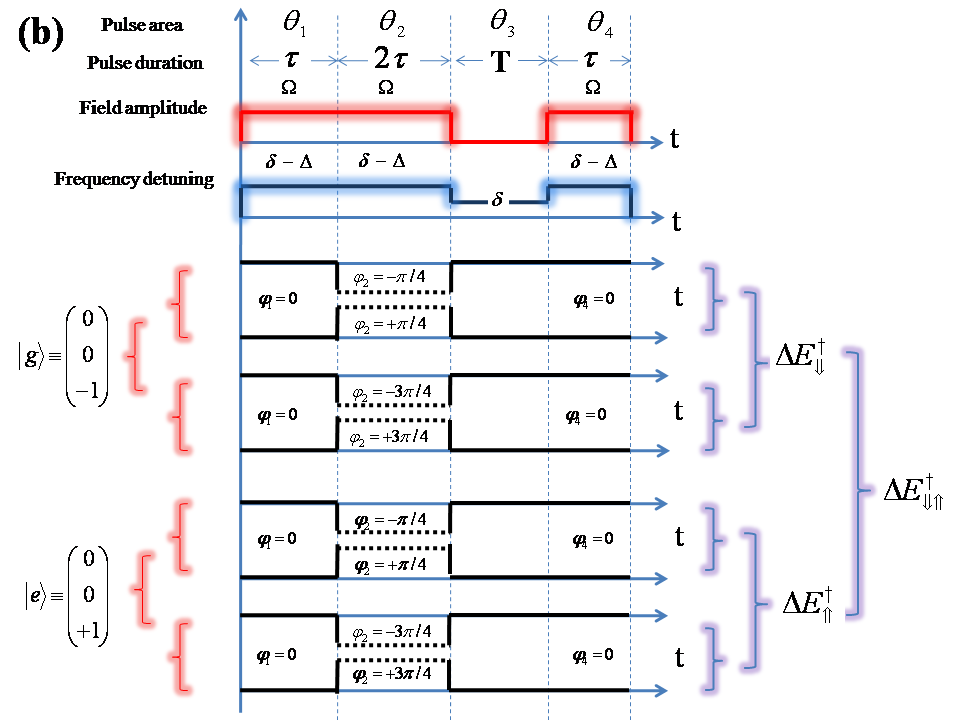}}
\end{minipage}
\hspace{0.8cm}
\begin{minipage}[c]{0.45\linewidth}
\centering
\resizebox{9.0cm}{!}{\includegraphics[angle=0]{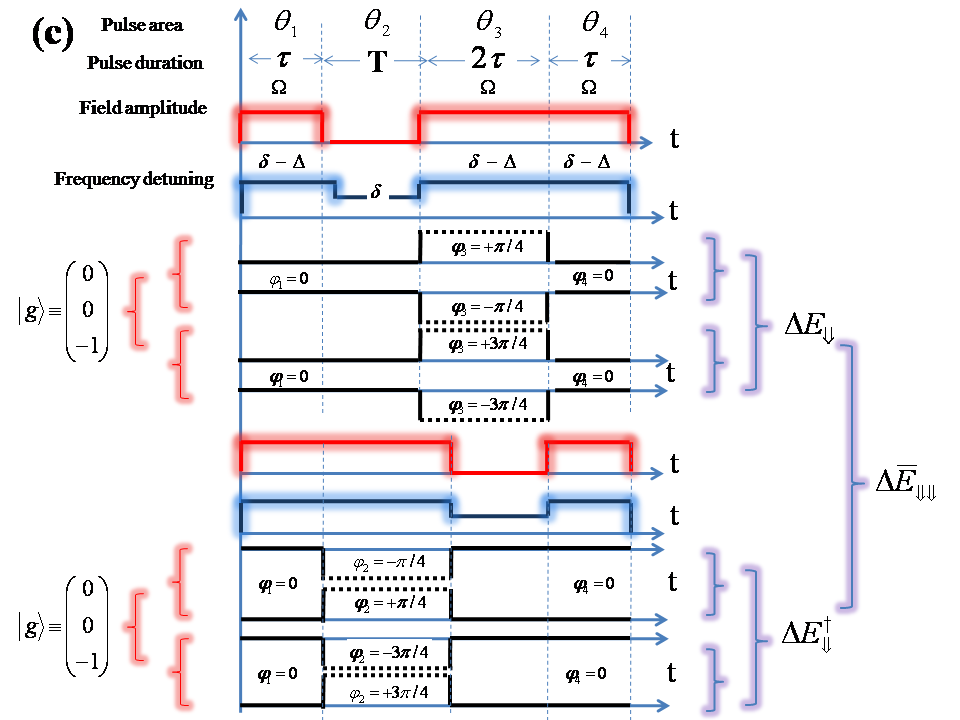}}
\end{minipage}
\caption{(color online) Universal laser frequency interrogation schemes for ultra robust frequency locking-points based on a combination of error signals generated by GHR($\pi/4$) and GHR($3\pi/4$) protocols from Table.~\ref{protocol-table-1}. (a) Interrogation protocol including a controllable population inversion between clock states. (b) Equivalent mirror-like interrogation protocol obtained by applying the transformation $t\rightarrow-t$ and $\varphi\rightarrow-\varphi$ on the scheme from (a). (c) Synthetic universal interrogation protocol by combining parts of (a) and (b) schemes, which eliminates population initialization in the upper state.}
\label{fig:GHR-protocol}
\end{figure*}
\indent We have also studied the influence of decoherence on Eq.~(\ref{eq:clock-frequency-shift}) with $\textup{GHR}(\pi/4)$ and $\textup{GHR}(3\pi/4)$ protocols
presented in Table.~\ref{protocol-table-1}. Clock frequency shift $\delta\widetilde{\nu}[\textup{GHR}(\pi/4)]$ and $\delta\widetilde{\nu}[\textup{GHR}(3\pi/4)]$ responses to laser parameter modifications are reported in 2D contour and density plot diagrams in Figs.~\ref{fig:2D-map-GHR-diagrams}(a1,a2) and (b1,b2).
It is worthwhile to note that if decoherence is vanishing, GHR$(\pi/4)$ and GHR$(3\pi/4)$ are indeed very efficient and lead to a complete suppression of probe-induced frequency shifts $\delta\widetilde{\nu}=0$ at all orders (see Table.~\ref{protocol-table-1}). It is why the figure equivalent to Figs.~\ref{fig:2D-map-HR-diagrams}(a1,a2) is not shown.
If the laser line-width is not negligible, generating decoherence $\gamma_{c}$, robustness of GHR protocols to laser power variation
and uncompensated frequency shifts are strongly degraded leading to Figs.~\ref{fig:2D-map-GHR-diagrams}. The radiative case $\Gamma\neq0$ leads to an important increase of clock frequency shifts $\delta\widetilde{\nu}[\textup{GHR}(\pi/4)]$ and $\delta\widetilde{\nu}[\textup{GHR}(3\pi/4)]$ and is not considered here.
\begin{figure*}[t!!]
\center
\resizebox{7.5cm}{!}{\includegraphics[angle=0]{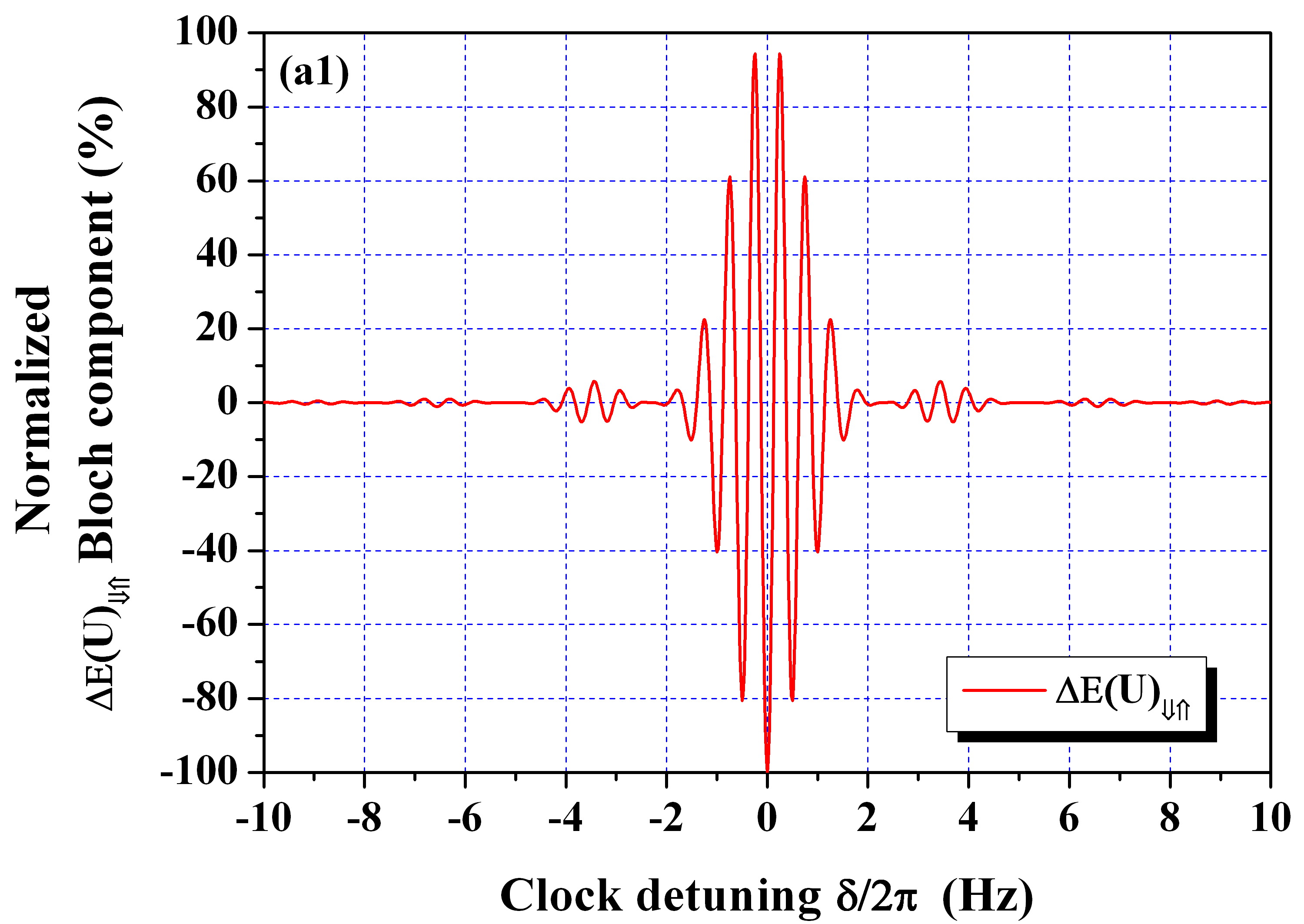}}\resizebox{7.5cm}{!}{\includegraphics[angle=0]{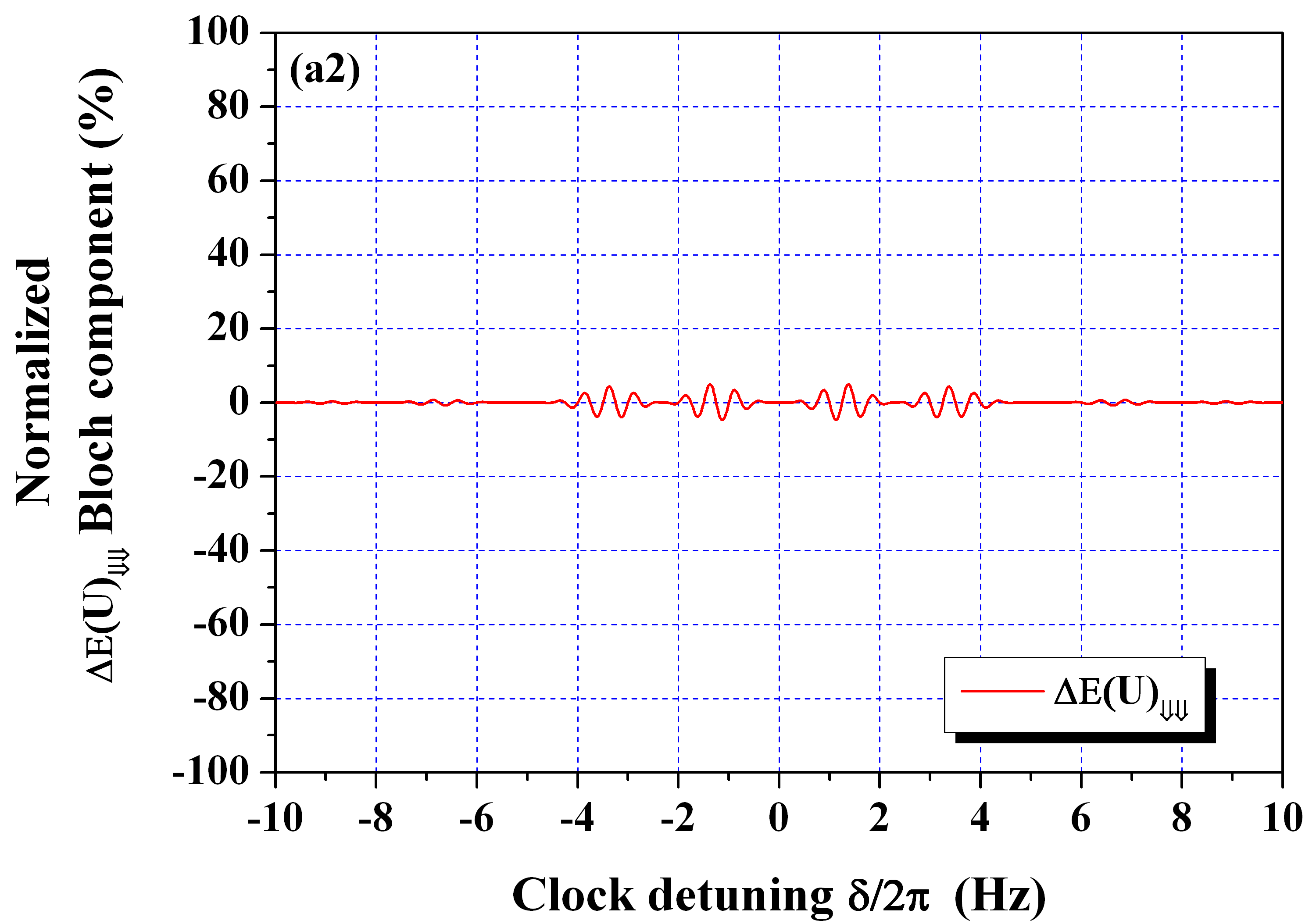}}
\resizebox{7.5cm}{!}{\includegraphics[angle=0]{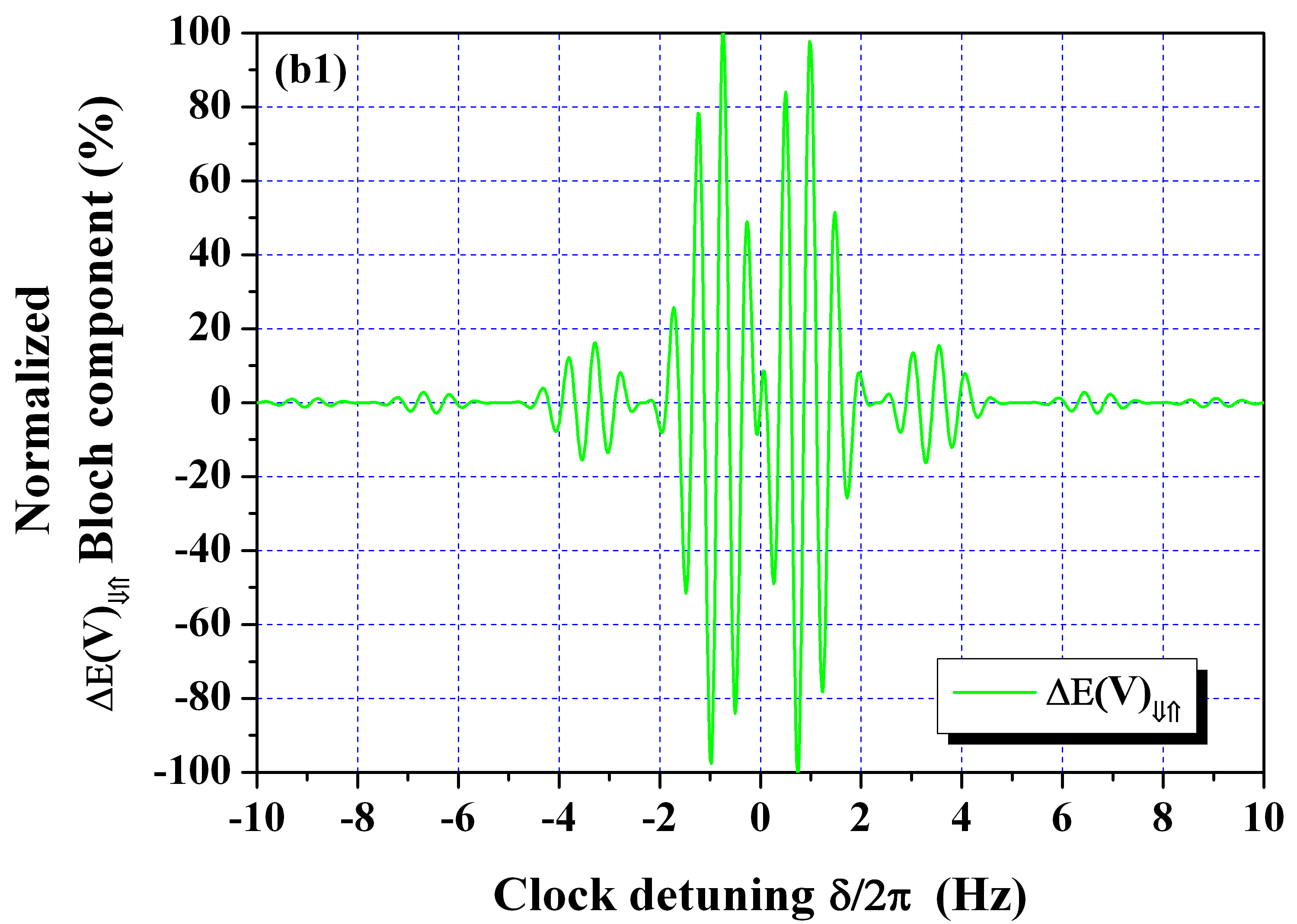}}\resizebox{7.5cm}{!}{\includegraphics[angle=0]{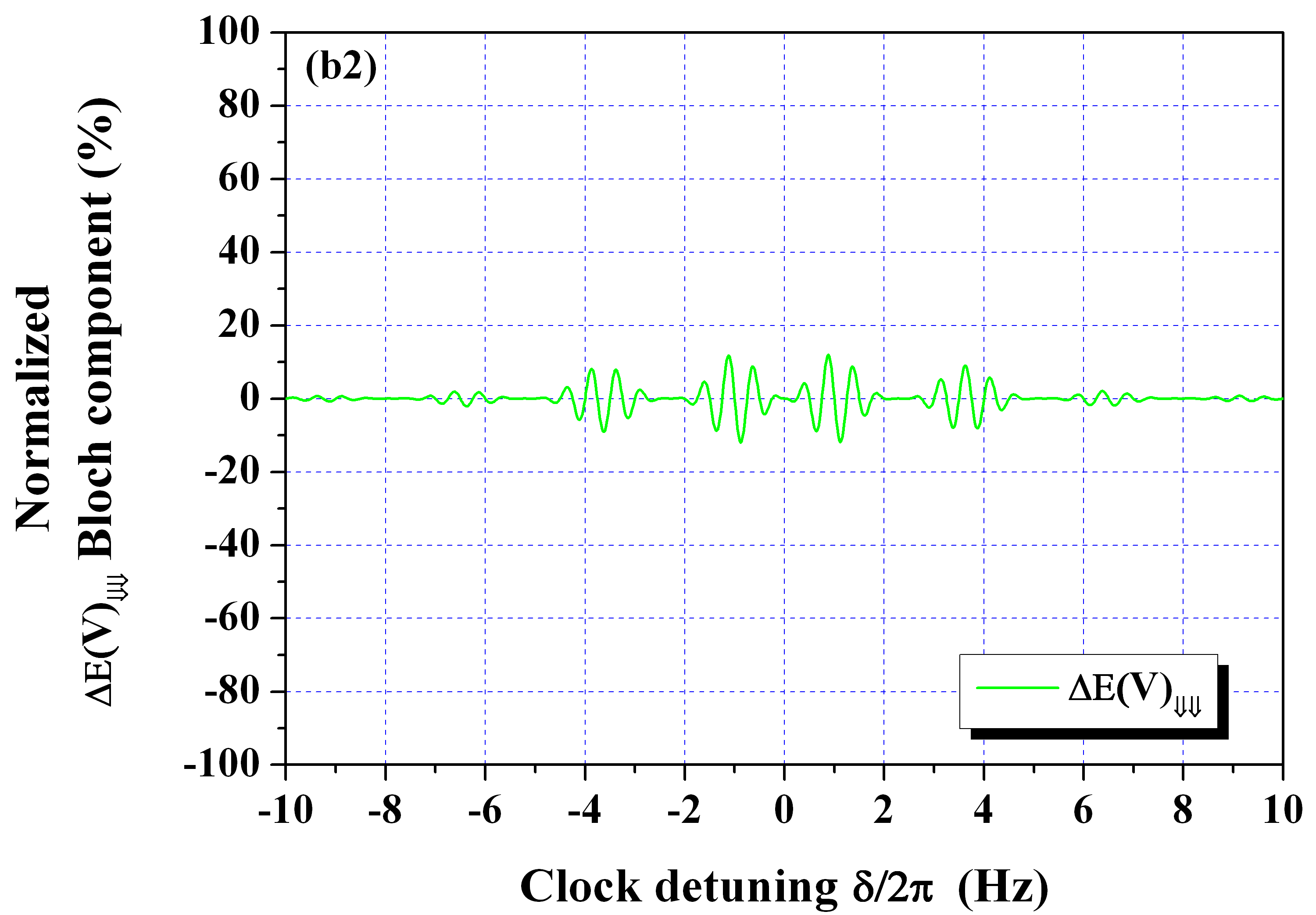}}
\resizebox{7.5cm}{!}{\includegraphics[angle=0]{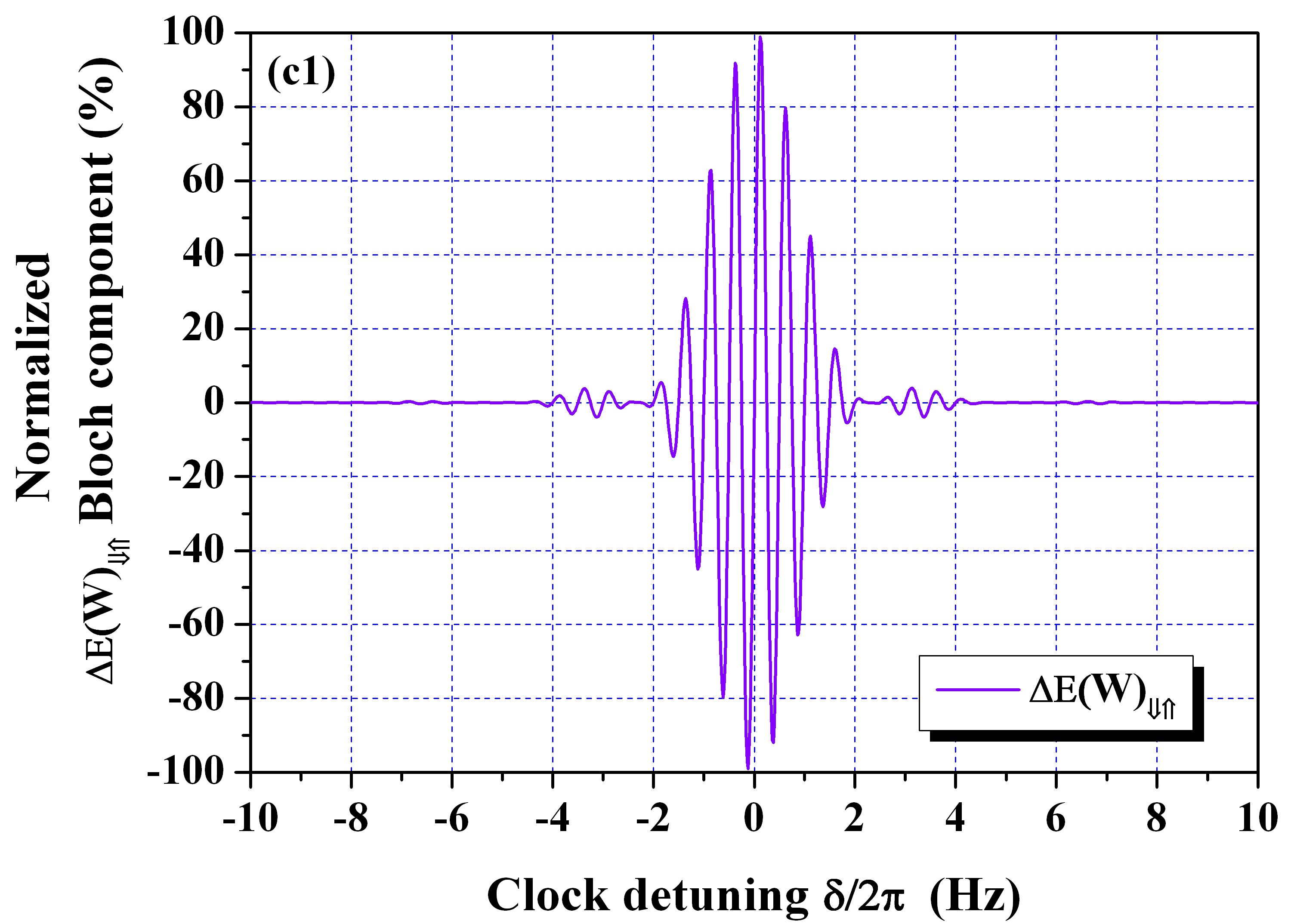}}\resizebox{7.5cm}{!}{\includegraphics[angle=0]{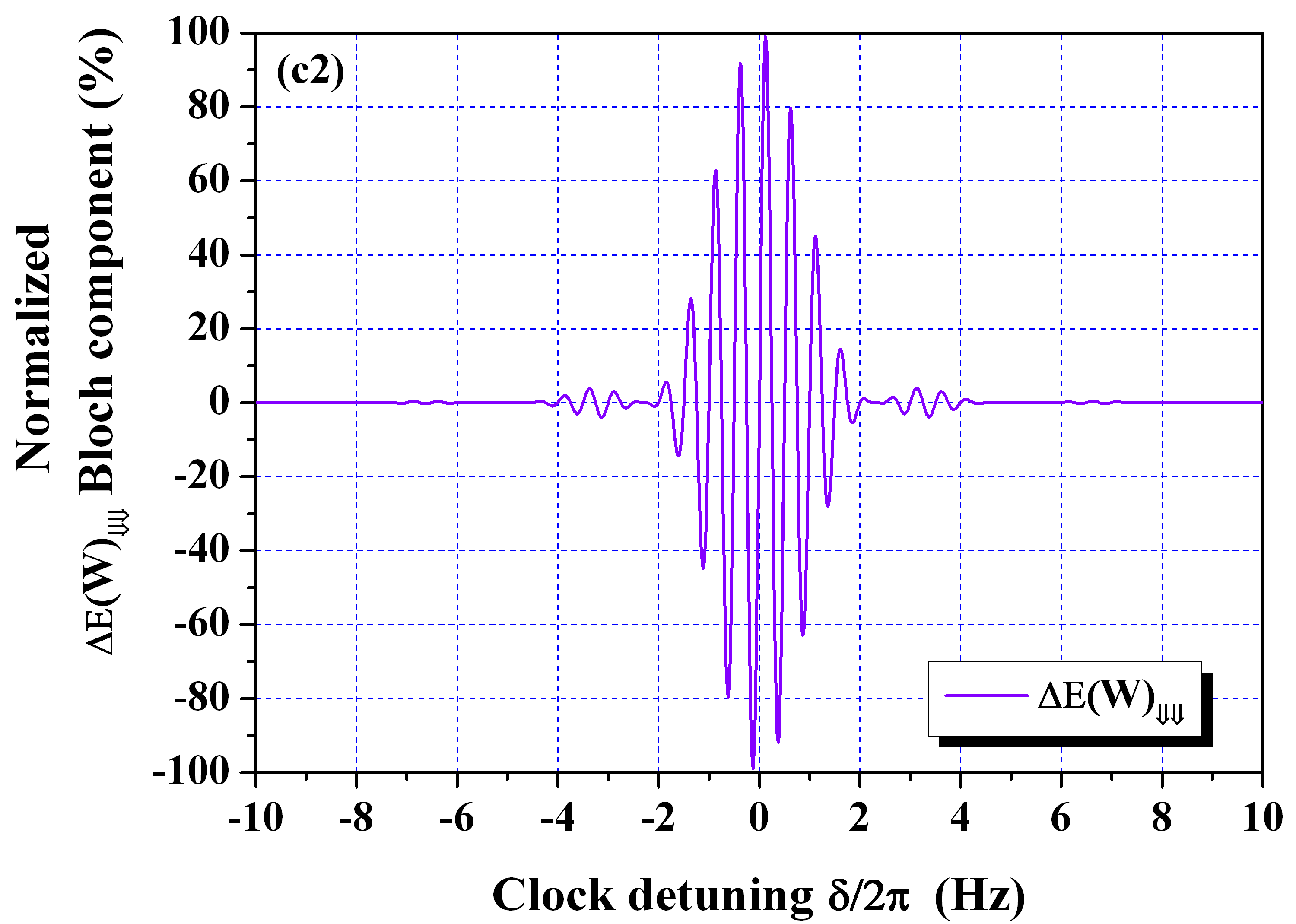}}
\caption{(color online) Comparison of normalized line-shapes and signal amplitudes of the three Bloch-vector component (U, V, W) versus clock frequency detuning $\delta/2\pi$ for protocols $\Delta\textup{E}_{\Downarrow\Uparrow}$ of Fig.~\ref{fig:GHR-protocol}(a,b)(left panels) and $\Delta\overline{\textup{E}}_{\Downarrow\Downarrow}$ of Fig.~\ref{fig:GHR-protocol}(c) (right panels). (a1) Bloch component $\Delta\textup{E(U)}_{\Downarrow\Uparrow}$ and (a2) $\Delta\overline{\textup{E}}\textup{(U)}_{\Downarrow\Downarrow}$. (b1) Bloch component $\Delta\textup{E(V)}_{\Downarrow\Uparrow}$ and (b2) $\Delta\overline{\textup{E}}\textup{(V)}_{\Downarrow\Downarrow}$. (c1) Bloch component $\Delta\textup{E}(\textup{W})_{\Downarrow\Uparrow}$ and (c2)
$\Delta\overline{\textup{E}}\textup{(W)}_{\Downarrow\Downarrow}$. The standard Rabi frequency for all pulses is $\Omega=\pi/2\tau$ where $\tau$ is the pulse duration reference. Pulse duration is $\tau=3/16$~s with free evolution time $\textup{T}=2$~s. We have ignored
probe-induced frequency shifts and dissipative processes for comparison between amplitude curves.}
\label{fig:universal-error-signal}
\end{figure*}
\begin{figure}[t!!]
\center
\resizebox{7.5cm}{!}{\includegraphics[angle=0]{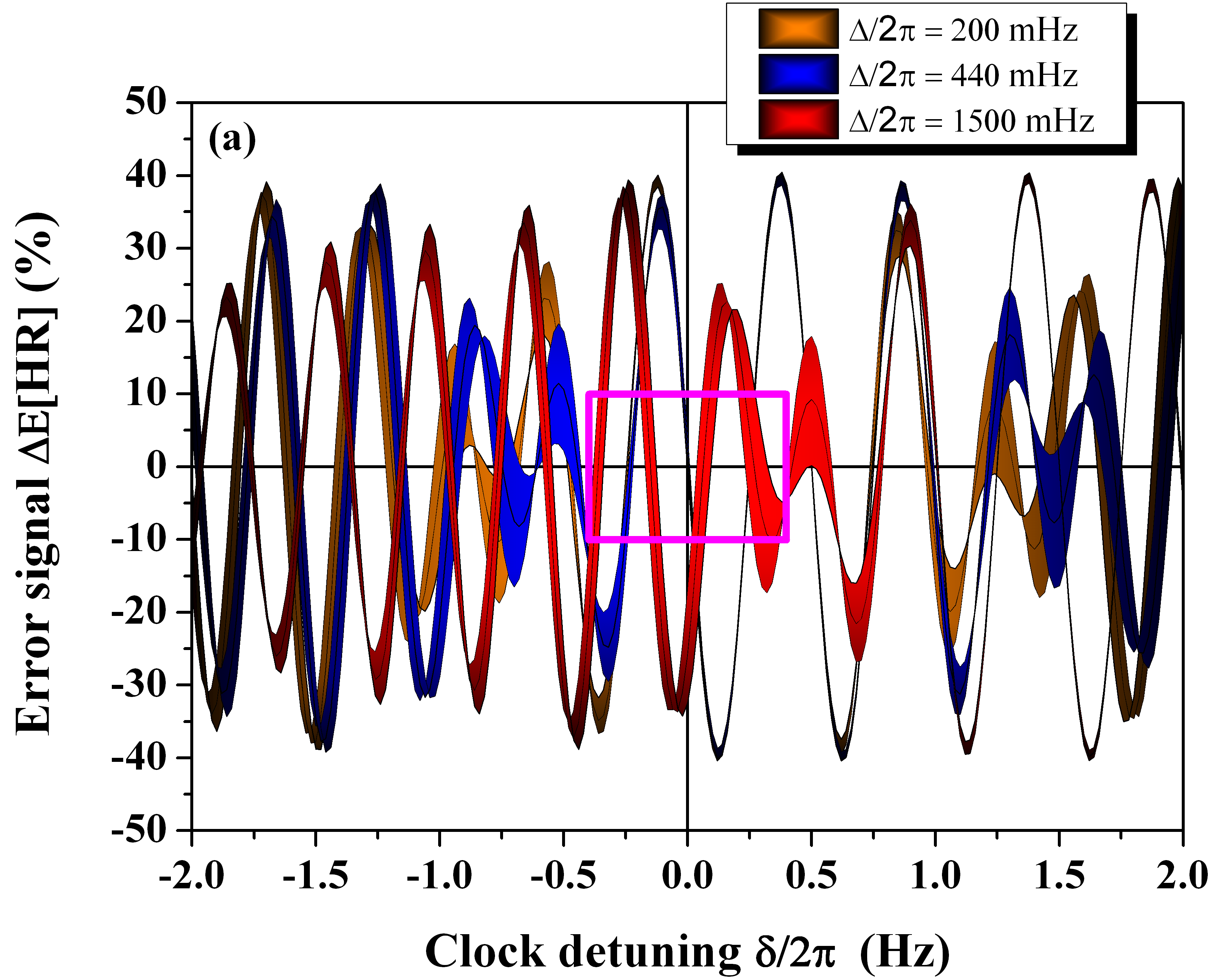}}
\resizebox{7.5cm}{!}{\includegraphics[angle=0]{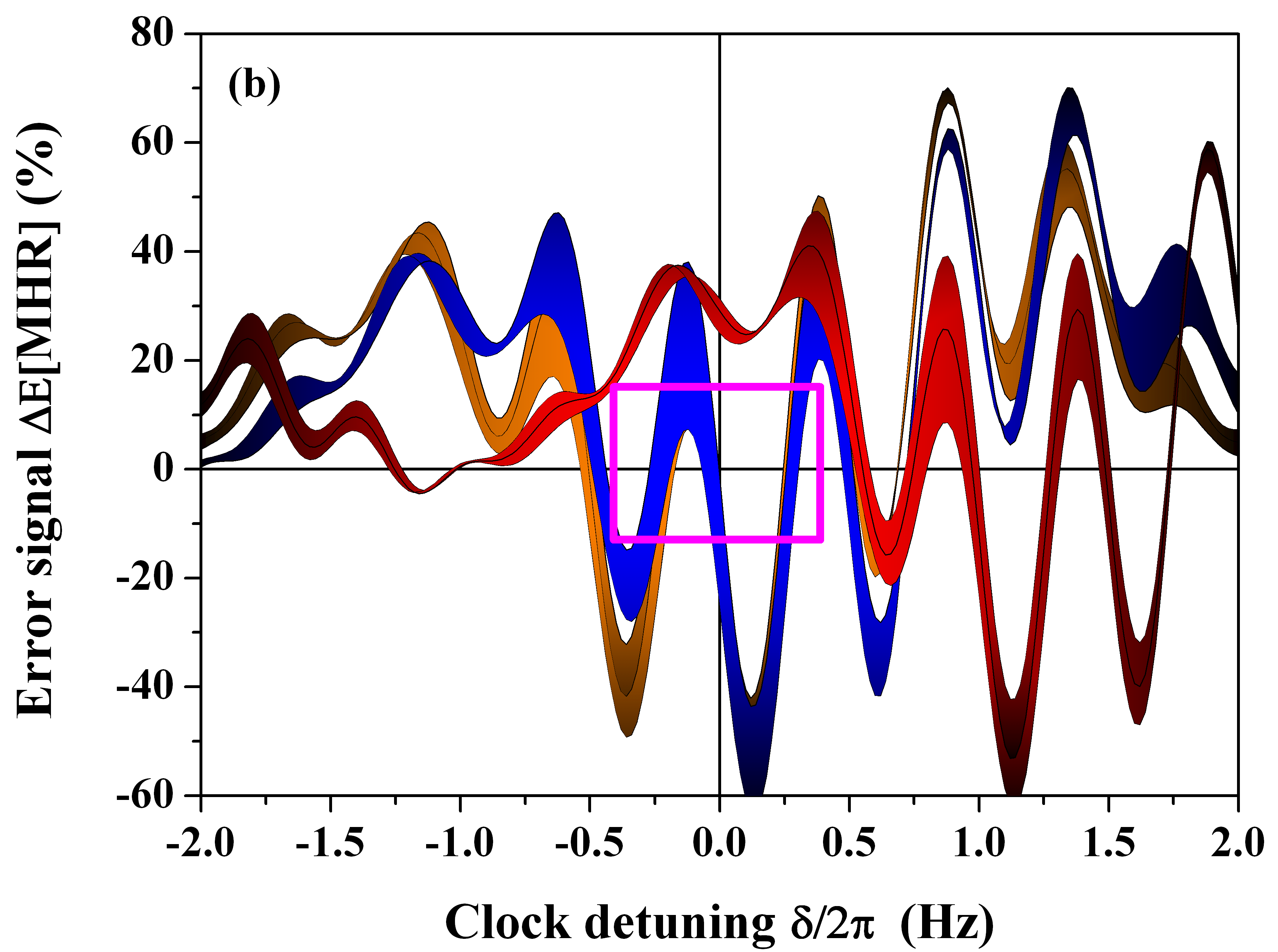}}
\resizebox{7.5cm}{!}{\includegraphics[angle=0]{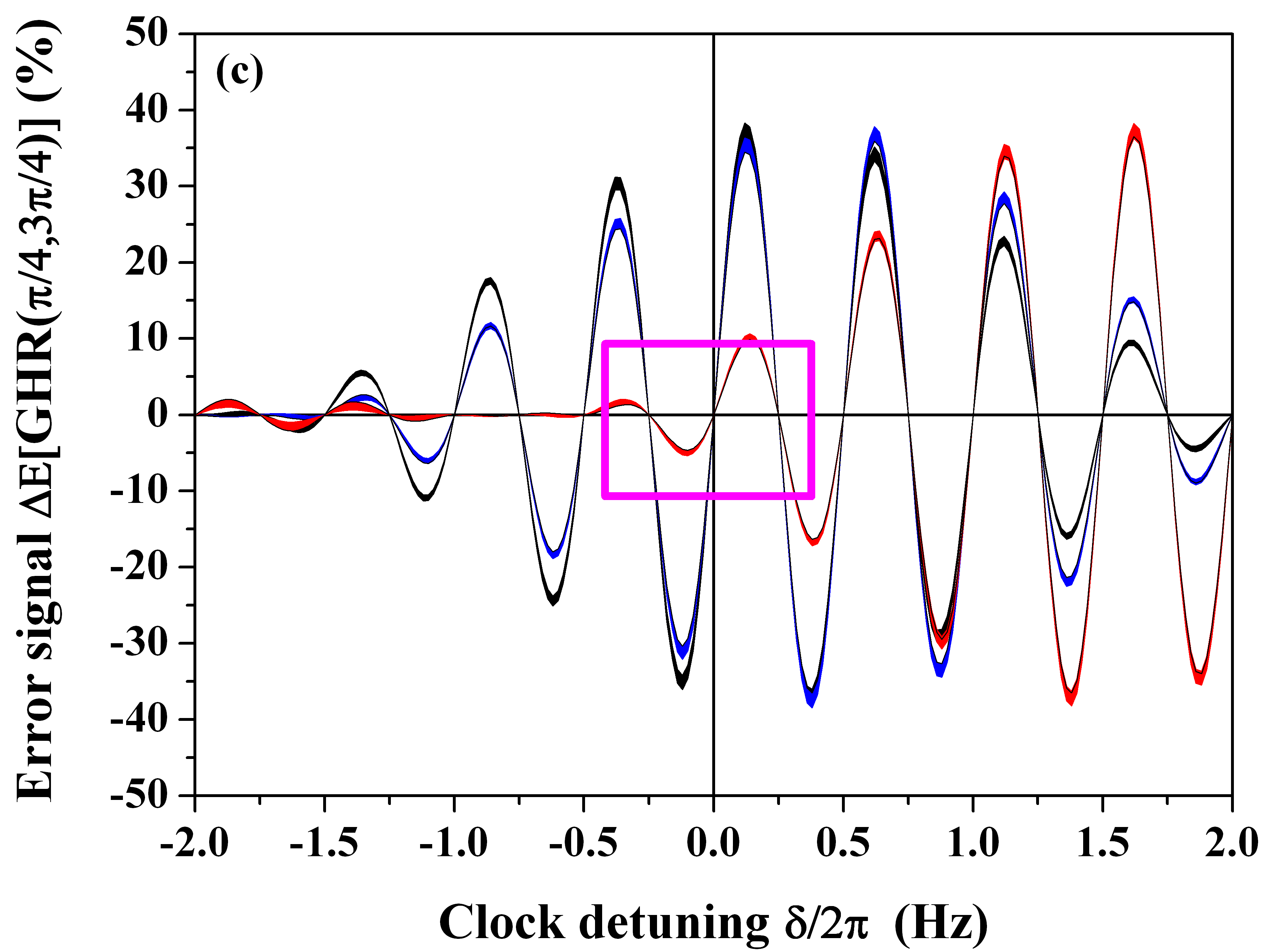}}
\caption{(color online) Error signal shapes versus clock frequency detuning $\delta/2\pi$ for three different uncompensated frequency shifts $\Delta/2\pi$. The laser frequency locking-point is delimited by a bounding box around $\delta\mapsto0$. (a) $\Delta\textup{E}[HR]$, (b) $\Delta\textup{E}[MHR]$ and (c) $\Delta\textup{E}_{\Downarrow\Uparrow}$.
Resilience of the frequency locking-point to various uncompensated probe-induced frequency shifts $\Delta/2\pi$ is demonstrated for the last scheme. Pulse area variation is set to $\Delta\theta/\theta=\pm10\%$ (shadow regions).
Dissipative parameters of the two-level system used as an atomic frequency reference are fixed to
$\gamma_{c}=2\pi\times50$~mHz, $\Gamma=2\pi\times100$~mHz and $\xi=0$. The standard Rabi frequency for all pulses is $\Omega=\pi/2\tau$ where $\tau$
is the pulse duration reference. Pulse duration is $\tau=3/16$~s with free evolution time $\textup{T}=2$~s.}
\label{fig:HR-GHR}
\end{figure}
The simultaneous observation of Figs.~\ref{fig:2D-map-GHR-diagrams}(a2) and (b2) shows that frequency locking-point regions of instability marked by different colored density plots are of opposite sign. It is thus possible to reconstruct another synthetic frequency-shift $\delta\widetilde{\nu}[\textup{syn}]$ to reliably suppress probe-induced shifts and their variations for $\textup{GHR}(\pi/4)$ and $\textup{GHR}(3\pi/4)$ interrogation schemes.
Taking the half-sum of the two clock frequency shifts $\delta\widetilde{\nu}[\textup{GHR}(\pi/4)]$ and
$\delta\widetilde{\nu}[\textup{GHR}(3\pi/4)]$ shown in Fig.~\ref{fig:2D-map-GHR-diagrams}(c1), (c2) displays small frequency locking-point stability islands near $\pi/2$ and $3\pi/2$ pulse area (pink and violet regions along the horizontal axis). This synthetic residual frequency-shift becomes much less sensitive to
variations in laser power and probe-shifts \cite{Yudin:2016}.

2D diagrams help in generating some stable regions by combining frequency-shift measurements when dissipative processes are present, but the process requires
a post-data treatment and the synthetic laser frequency locking-point is never absolutely protected against residual probe-shifts and laser power variations degrading the clock stability.

\section{UNIVERSAL ELIMINATION PROTOCOL OF PROBE-FIELD-INDUCED FREQUENCY SHIFTS}

\indent Although a recent frequency-shift extrapolation-based method based on multiple (HR) schemes with different free evolution times was
proposed to reduce imperfect compensation of probe-induced shifts well below a fractional frequency change of $10^{-18}$
\cite{Yudin:2016}, the existence of an absolute interrogation protocol directly canceling these shifts on the dispersive error signal shape at all orders even in presence of decoherence, relaxation by spontaneous emission and collisions was not yet established.

To solve the problem, we propose in Fig.~\ref{fig:GHR-protocol}
a universal interrogation protocol denoted GHR$(\pi/4,3\pi/4)$ based on mixing GHR$(\pi/4)$ and GHR$(3\pi/4)$ schemes
from Table.~\ref{protocol-table-1}, interleaved or not by a controllable population inversion between clock states.
Symmetric properties of the new interrogation scheme might be even exploited with
some quantum logic gate circuits using entanglement of prepared qubits \cite{Tan:2015}, reducing the
number of measurements required to generate the correct laser frequency locking-point.
Unlike the synthetic frequency-shift realization presented in Fig.~\ref{fig:2D-map-GHR-diagrams}(c2) which requires to combine two separated clock frequency shift evaluations potentially degrading the clock stability, our new universal protocol generates a direct laser frequency-locking point as a strong error signal gradient robust even to substantial uncompensated frequency shifts.

The universal interrogation protocol $\textup{GHR}(\pi/4,3\pi/4)$ breaks down into three different layouts of composite optical-pulses as shown in Fig.~\ref{fig:GHR-protocol}(a,b,c). The initial combination of GHR($\pi/4$) and GHR($3\pi/4$) protocols from Table.~\ref{protocol-table-1}, presented in Fig.~\ref{fig:GHR-protocol}(a), includes a population inversion between clock states.
Phase-steps are applied only during the third pulse interaction following a free evolution time. A similar interrogation scheme can be realized
using reverse composite pulses as in Fig.~\ref{fig:GHR-protocol}(b) with mirror-like protocols denoted by $\dagger$-type from Table.~\ref{protocol-table-1}.
In such a case, while ignoring stationary-states, a time and phase reversal symmetry transformation can be applied on the scheme presented in Fig.~\ref{fig:GHR-protocol}(a) to recover an identical line-shape obtained with Fig.~\ref{fig:GHR-protocol}(b) and mirror-like protocol \cite{Zanon:2015}.
A new frequency locking-point can still be synthesized as shown in Fig.~\ref{fig:GHR-protocol}(c) mixing some parts of the two previous protocols
while eliminating population initialization in the upper state. Such an alternative scheme might be seen as a sort of spin echoe hybrid technique \cite{Hahn:1950} removing some uncontrollable variations of laser parameters with time order pulse reversal.
In all cases, the new error signal $\Delta\textup{E}[\textup{GHR}(\pi/4,3\pi/4)]$ requires a specific number of atomic population fraction measurements to generate a robust laser frequency locking-point depending on the nature of the dissipative processes impacting the atomic transition. The ideal laser frequency-locking point with no correction for uncompensated probe induced-frequency shifts is provided by the use of $\pm\pi/4$ and $\pm3\pi/4$ phase-steps which are canceling exactly steady-state solutions from Bloch solutions (see for example Eq.~(\ref{eq:combination}) in appendix D).

When an ideal two-level system is considered, error signals based on MHR and GHR protocols require only 2 population fraction measurements generating a very
stable frequency locking-point with full elimination of residual clock frequency shifts $\delta\widetilde{\nu}$ as reported in Table.~\ref{protocol-table-1}.
For a pure decoherence case affecting the frequency locking-point stability as shown in Fig.~\ref{fig:2D-map-GHR-diagrams}(a2) and Fig.~\ref{fig:2D-map-GHR-diagrams}(b2),
a combination of 4 atomic population fraction measurements with $\pm\pi/4$ and $\pm3\pi/4$ phase-steps and one single state initialization (half-part of the universal protocol
from Fig.~\ref{fig:GHR-protocol}(a) or Fig.~\ref{fig:GHR-protocol}(b) called $\dagger$-type), is sufficient to totally cancel the probe-induced frequency shifts.
The normalized error signal denoted $\Delta\textup{E}_{\Downarrow(\Uparrow)}\equiv\Delta\textup{E}[\textup{GHR}(\pi/4,3\pi/4)]$ (or equivalently $\dagger$-type) is generated as follows:
\begin{equation}
\begin{split}
\Delta\textup{E}_{\Downarrow(\Uparrow)}&=\frac{1}{2}\left(\Delta\textup{E}[\textup{GHR}(\pi/4)]-\Delta\textup{E}[\textup{GHR}(3\pi/4)]\right)_{\Downarrow(\Uparrow)},\\
\Delta\textup{E}^{\dagger}_{\Downarrow(\Uparrow)}&=\frac{1}{2}\left(\Delta\textup{E}^{\dagger}[\textup{GHR}(\pi/4)]-\Delta\textup{E}^{\dagger}[\textup{GHR}(3\pi/4)]\right)_{\Downarrow(\Uparrow)}.
\end{split}
\label{eq:error-signal-decoherence}
\end{equation}
where $\Downarrow(\Uparrow)$ means the protocol is applied with population initialization in either ground state $|g\rangle\equiv\Downarrow$ or excited state $|e\rangle\equiv\Uparrow$.

For simultaneous activation of spontaneous emission and decoherence, the error signal thus requires
8 atomic population measurements divided into 4 measurements with state initialization in $|g\rangle$ and $|e\rangle$ (see Fig.~\ref{fig:GHR-protocol}(a)).
The dispersive error signal $\Delta\textup{E}_{\Downarrow\Uparrow}\equiv\Delta\textup{E}[\textup{GHR}(\pi/4,3\pi/4)]$ ($\Delta\textup{E}^{\dagger}_{\Downarrow\Uparrow}$) now becomes:
\begin{equation}
\begin{split}
\Delta\textup{E}_{\Downarrow\Uparrow}=\frac{1}{2}\left(\Delta\textup{E}_{\Downarrow}-\Delta\textup{E}_{\Uparrow}\right)=
\textup{F}_{\Downarrow\Uparrow}\left[M(0),\theta_{1},\theta_{3}\right]\sin(\delta\textup{T}),\\
\Delta\textup{E}^{\dagger}_{\Downarrow\Uparrow}=\frac{1}{2}\left(\Delta\textup{E}^{\dagger}_{\Downarrow}-\Delta\textup{E}^{\dagger}_{\Uparrow}\right)=
\textup{F}^{\dagger}_{\Downarrow\Uparrow}\left[M(0),\theta_{1},\theta_{3}\right]\sin(\delta\textup{T}).
\end{split}
\label{eq:error-signal-relaxation-decoherence}
\end{equation}
where amplitude functions $\textup{F}\left[M(0),\theta_{1},\theta_{3}\right]$ can be derived from appendix D.

Note that an additional protocol presented in Fig.~\ref{fig:GHR-protocol}(c) can synthesize another
ultra stable frequency locking-point while avoiding population initialization in both quantum states. We apply now a linear combination of error signals
from two opposite sequence of composite laser-pulses that are reversed in time ordering as:
\begin{equation}
\begin{split}
\Delta\overline{\textup{E}}_{\Downarrow\Downarrow}=\frac{1}{4}\left(\Delta\textup{E}^{\dagger}_{\Downarrow}+\Delta\textup{E}_{\Downarrow}\right)=
F_{\Downarrow\Downarrow}\left[M(0),\theta_{1},\theta_{3}\right]\sin(\delta\textup{T}).
\end{split}
\label{eq:error-signal-relaxation-decoherence-reversion}
\end{equation}
where amplitude function $\textup{F}\left[M(0),\theta_{1},\theta_{3}\right]$ can be derived from appendix E.
Note that such an error signal $\Delta\overline{\textup{E}}_{\Downarrow\Downarrow}$ is remarkable over a few additional features. It produces a zero crossing point with enhanced immunity to residual offset variations independent of a perfect quantum state initialization.\\
\begin{table*}[t!!]
\centering%
\caption{Absolute robustness of various error signal laser frequency locking-points to individual or multiple $\{\}$ dissipative parameters $\gamma_{c},\Gamma,\xi$ for a closed two-level system. The number of atomic state population measurements N required to build the error signal is also indicated. A perfect phase-stepping of the laser for all protocols is considered here.
Note if $\Gamma\neq0$, then $\gamma_{c}=\Gamma/2$ to be consistent with a pure radiative process.}
\label{protocol-table-2}
\renewcommand{\arraystretch}{1.4}
\begin{tabular}{|c|c|ccccc|}
\hline
\hline
 & & & & & & \\
 error signal & N & $\gamma_{c}$ &  $\xi$ &  $\{\gamma_{c},\xi\}$ & $\{\gamma_{c},\Gamma\}$ & $\{\gamma_{c},\Gamma,\xi\}$ \\
\hline
 \begin{tabular}{cc}
   $\Delta\textup{E}[\textup{HR}]$ & $\Delta\textup{E}[\textup{GHR}(\pi/4)]$ \\
    $\Delta\textup{E}[\textup{MHR}]$  &  $\Delta\textup{E}[\textup{GHR}(3\pi/4)]$ \\
 \end{tabular}
 & 2 & NO &  NO  & NO & NO & NO \\
\hline
\hline
   \begin{tabular}{c} $\Delta\textup{E}_{\Downarrow(\Uparrow)}$, $\Delta\textup{E}^{\dagger}_{\Downarrow(\Uparrow)}$
    \end{tabular}
     & 4  & \checkmark & \checkmark & \checkmark  &  NO  &  NO \\
\hline
\hline
 \begin{tabular}{c}   $\Delta\textup{E}_{\Downarrow\Uparrow}$, $\Delta\textup{E}^{\dagger}_{\Downarrow\Uparrow}$,
  $\Delta\overline{\textup{E}}_{\Downarrow\Downarrow}$
 \end{tabular} & 8 & \checkmark & \checkmark & \checkmark &  \checkmark & \checkmark \\
\hline
\hline
\end{tabular}
\end{table*}
We have reported all Bloch-vector component error signal line-shapes with normalized amplitudes for $\Delta\textup{E}_{\Downarrow\Uparrow}$ and
$\Delta\textup{E}_{\Downarrow\Downarrow}$ in Fig.~(\ref{fig:universal-error-signal}) ignoring dissipative processes. Curves from the right panels are normalized respectively to the ones from left panels showing very different signal strengths in real and imaginary parts of the optical coherence under identical choice of laser parameters. Notice that if a simultaneous laser probe transmission monitoring is allowed with the first universal protocol $\Delta\textup{E}_{\Downarrow\Uparrow}$, a parallel implementation of a feed-back loop control may be realized by recording
imaginary and real parts of the optical coherence (see Eq.~\ref{eq:final-error-signal} in appendix D and Fig.~\ref{fig:universal-error-signal}(a1,b1)) as an additional hint signal to steer any probe frequency-drift in the correct direction over long periods of time.
\begin{table}[b!!]
\centering%
\caption{Estimation of various laser frequency locking-point frequency shifts $|\delta\widetilde{\nu}|$ (absolute value) to a systematic error in laser phase-steps by $\delta\varphi/\varphi=\pm1\%$. Dissipative atomic parameters are fixed to $\Gamma=2\pi\times100$~mHz and $\gamma_{c}=\Gamma/2$ under pulse area variation by $\Delta\theta/\theta=\pm10\%$ for three residual uncompensated frequency shifts $\Delta/2\pi$. All frequency shifts are in given in mHz unit.}
\label{protocol-table-3}
\renewcommand{\arraystretch}{1.2}
\begin{tabular}{|c|c|c|c|}
\hline
\hline
 \begin{tabular}{c} error signal \end{tabular} & \begin{tabular}{c} $\Delta/2\pi=0$  \end{tabular} & \begin{tabular}{c} $\Delta/2\pi=440$ \end{tabular}  & \begin{tabular}{c} $\Delta/2\pi=1500$   \end{tabular} \\
\hline
\begin{tabular}{c} $\Delta\textup{E}[\textup{MHR}]$ \end{tabular} & $|\delta\widetilde{\nu}|\leq100$ & $|\delta\widetilde{\nu}|\leq100$ & $|\delta\widetilde{\nu}|\leq1000$ \\
\hline
\hline
 \begin{tabular}{c} $\Delta\textup{E}[\textup{HR}]$ \end{tabular} & $|\delta\widetilde{\nu}|\leq10$ & $|\delta\widetilde{\nu}|\leq10$ & out of range \\
\hline
\hline
\begin{tabular}{c} $\Delta\overline{\textup{E}}_{\Downarrow\Downarrow}$ \end{tabular}& $|\delta\widetilde{\nu}|=0$ & $|\delta\widetilde{\nu}|\leq1$ & $|\delta\widetilde{\nu}|\leq10$ \\
\hline
\hline
\begin{tabular}{c} $\Delta\textup{E}_{\Downarrow\Uparrow}$ \end{tabular} & $|\delta\widetilde{\nu}|=0$ & $|\delta\widetilde{\nu}|\leq0.1$ & $|\delta\widetilde{\nu}|\leq1$ \\
\hline
\hline
\end{tabular}
\end{table}
The second universal protocol $\Delta\textup{E}_{\Downarrow\Downarrow}$ does even not require any elimination of residual optical coherence after the entire interrogation process because non vanishing real and imaginary parts of any optical coherence, which may interfere with the laser probe during the pulse spectroscopy \cite{Prestage:2010}, are exhibiting the same dispersive line-shape locked at the unperturbed clock frequency (see Eq.~(\ref{eq:combination-error-signal}) in appendix E).\\
\begin{figure}[t!!]
\center
\resizebox{7.5cm}{!}{\includegraphics[angle=0]{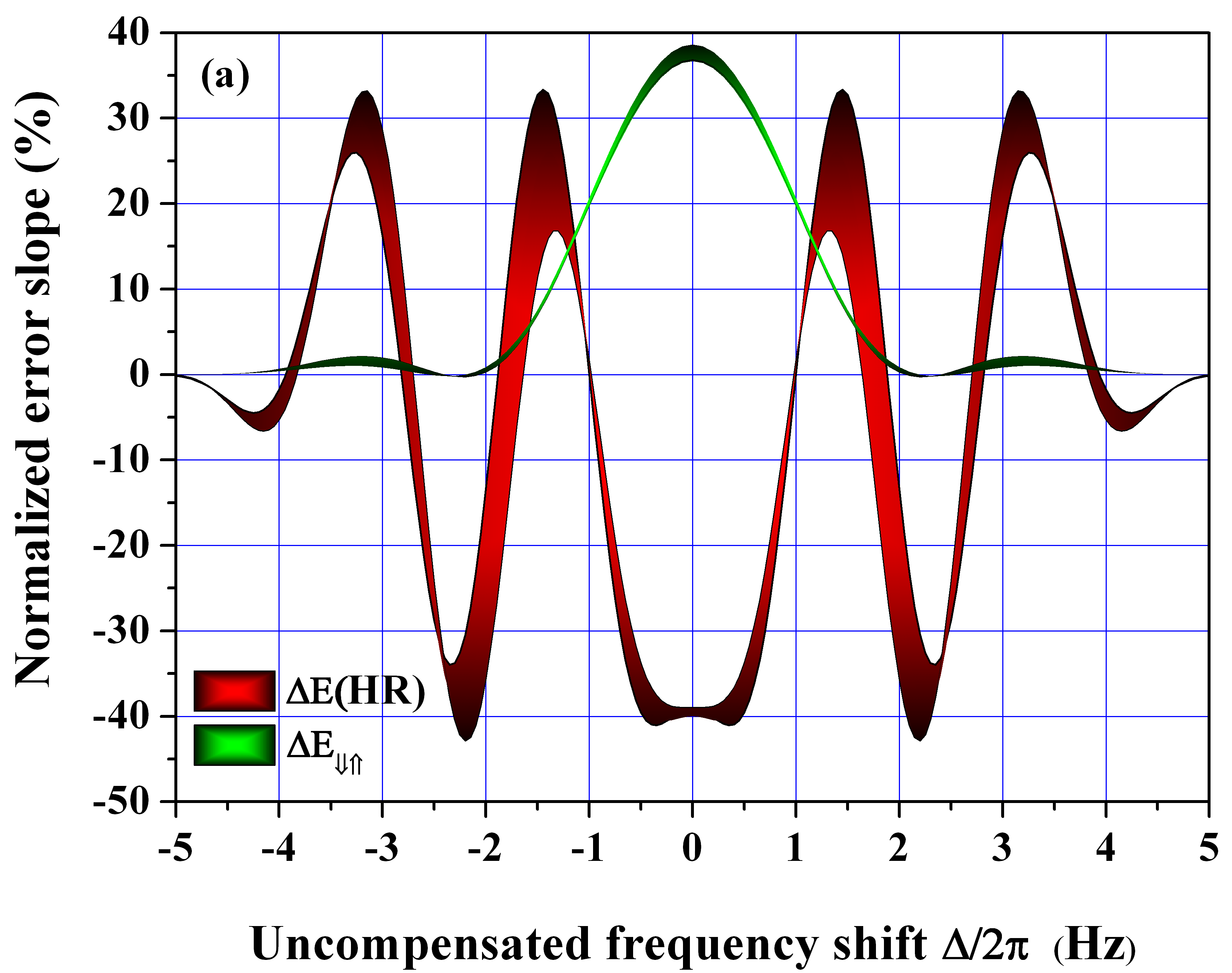}}
\resizebox{7.5cm}{!}{\includegraphics[angle=0]{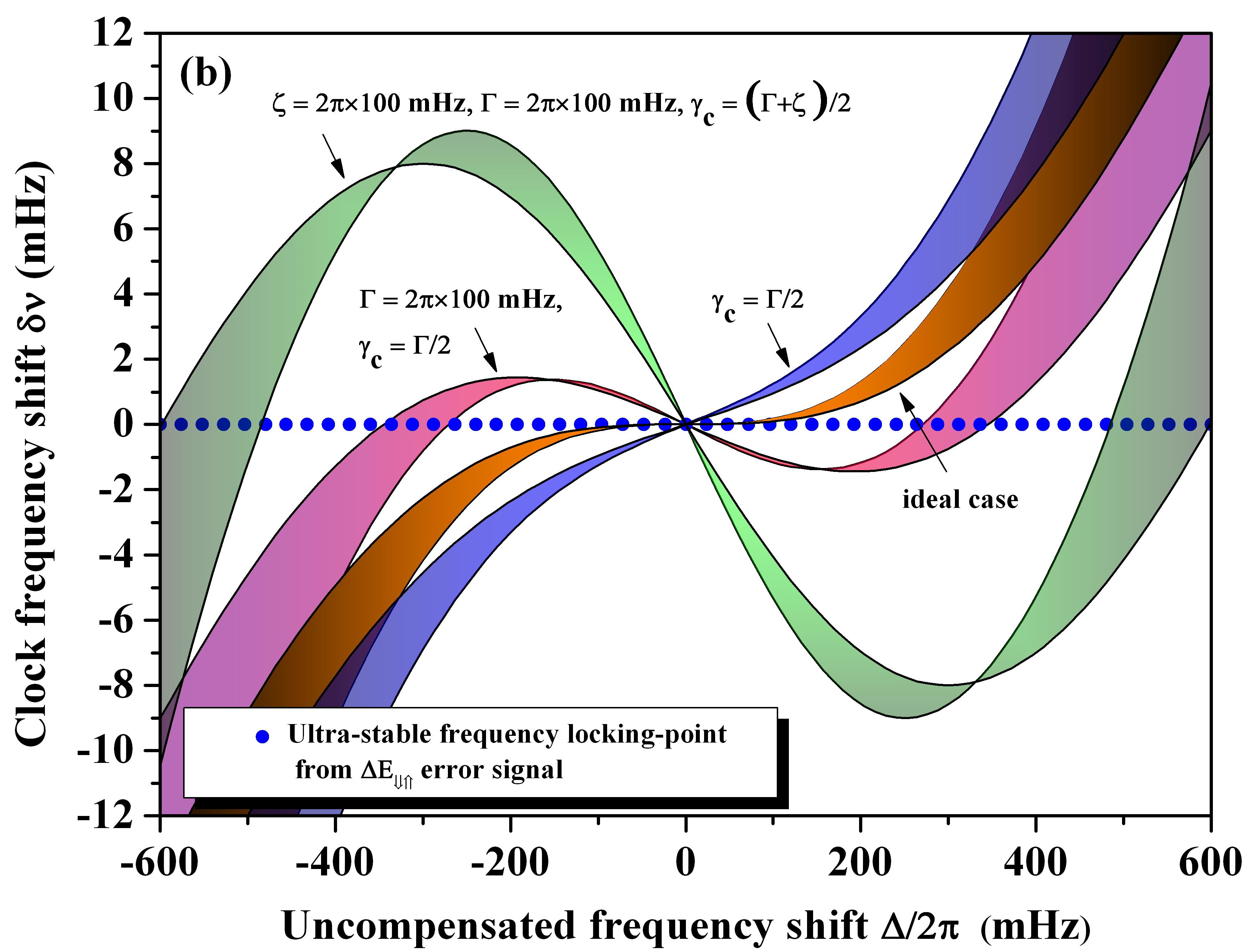}}
\caption{(color online) Compared robustness of $\Delta\textup{E}[\textup{HR}]$ and $\Delta\textup{E}_{\Downarrow\Uparrow}$ schemes with $\Delta\theta/\theta=\pm10\%$ error in the laser field amplitude. (a) Acceptance bandwidth of normalized $\Delta\textup{E}[\textup{HR}]$ and $\Delta\textup{E}_{\Downarrow\Uparrow}$ ($\Delta\textup{E}^{\dagger}_{\Downarrow\Uparrow}$) error signal
slopes versus uncompensated probe-shifts $\Delta/2\pi$. (b) $\Delta\textup{E}_{\Downarrow\Uparrow}$ (blue dots) and $\Delta\textup{E}[\textup{HR}]$ (solid line) clock frequency shifts when decoherence, relaxation by spontaneous emission and collisions are toggled on-off. All other parameters are identical to Fig.~\ref{fig:HR-GHR}.}
\label{fig:GHR-HR-clock-frequency-shifts}
\end{figure}
We have respectively reported in Fig.~\ref{fig:HR-GHR}(a),(b),(c) typical error signal patterns $\Delta\textup{E}[HR]$, $\Delta\textup{E}[MHR]$ and $\Delta\textup{E}_{\Downarrow\Uparrow}$ versus the clock detuning under simultaneous action of decoherence and relaxation. It is clearly demonstrated that the laser frequency locking-point generated by $\Delta\textup{E}_{\Downarrow\Uparrow}$ is ultra-robust against dissipation and residual uncompensated probe-induced frequency shifts compared to other schemes.
The robustness of the normalized error signal slope to uncompensated frequency shifts and pulse area variation is presented in Fig.~\ref{fig:GHR-HR-clock-frequency-shifts}(a).
The $\Delta\textup{E}_{\Downarrow\Uparrow}$ ($\Delta\textup{E}^{\dagger}_{\Downarrow\Uparrow}$) acceptance bandwidth is two times larger than the $\Delta\textup{E}[\textup{HR}]$ error signal under identical laser parameters. A very large $\pm10\%$ error on the laser field amplitude slightly modifies the slope but does not degrade
the frequency locking range where the slope does not drop to zero. We have also checked that all universal interrogation schemes do not need to rely on a perfect 100$\%$ initialization of the excited state. Investigating various analytical $\Delta\textup{E}_{\Downarrow\Uparrow}$ ($\Delta\textup{E}^{\dagger}_{\Downarrow\Uparrow}$) error signal shape expressions (see appendix D), non ideal population inversion between quantum states, for example due to large frequency shifts induced by a $\pi$ pulse excitation, will lead only to a linear reduction in size amplitude of the generated error signal with no deterioration of the laser frequency locking-point robustness.

We finally report in Fig.~\ref{fig:GHR-HR-clock-frequency-shifts}(b) the sensitivity of $\delta\widetilde{\nu}[\textup{HR}]$ and
$\delta\widetilde{\nu}[\Delta\textup{E}_{\Downarrow\Uparrow}]$ clock frequency shifts to residual uncompensated probe-shifts for pulse area variations of $\Delta\theta/\theta=\pm10\%$ and various dissipative processes configurations already displayed.
The $\delta\widetilde{\nu}[\textup{HR}]$ clock frequency shift measurement
affecting the central fringe minimum for the HR protocol is shown in Fig.~\ref{fig:GHR-HR-clock-frequency-shifts}(b).
It is worth to note the perfect cancelation of the locked frequency shift $\delta\widetilde{\nu}[\Delta\textup{E}_{\Downarrow\Uparrow}]$ reported as blue dots, even in presence of decoherence and relaxation.
Table.~\ref{protocol-table-2} summarizes absolute robustness of different error signal laser frequency locking-points to various combination of dissipative parameters $\gamma_{c},\Gamma,\xi$, for a closed two-level system.
Table.~\ref{protocol-table-3} reports the ultimate clock frequency shift sensitivity from different error signal laser frequency locking-points to a systematic error in laser phase-stepping process by $\delta\varphi/\varphi=\pm1\%$ under a strong pulse area variation by $\Delta\theta/\theta=\pm10\%$ and for three different values of uncompensated frequency-shifts. From this analysis, the $\Delta\textup{E}[\textup{MHR}]$ error signal presents a systematic parasitic shift even when a complete elimination of residual frequency-shifts $\Delta$ is realized. The result is also consistent with a previous numerical work which was only for a pure decoherence effect \cite{Yudin:2016}. Based on our complete GHR lineshape solution, we have been able to identify that the $\Delta\textup{E}[\textup{HR}]$ error signal also recovers a small parasitic shift due to a small imperfect phase-step modulation.
We note that for uncompensated frequency shifts larger than $\Delta/2\pi\geq1$~Hz, the HR slope has abruptly changed in sign as expected from Fig.~\ref{fig:GHR-HR-clock-frequency-shifts}(a) leading to a rapid loss of clock stability. However the larger acceptance bandwidth for universal schemes $\Delta\overline{\textup{E}}_{\Downarrow\Downarrow}$ and $\Delta\textup{E}_{\Downarrow\Uparrow}$ allows for maintaining a robust error signal gradient even for larger uncompensated probe-induced frequency-shifts. They are still ultimately suffering from a weak linear dependance to these uncompensated frequency-shifts because a small error in phase-steps break the perfect rejection of steady-state solutions when combining several atomic population fraction measurements.

\section*{CONCLUSIONS}

\indent We have first established the analytical expression of the Bloch vector evolution in presence of decoherence and relaxation during a light pulse interaction. We then deduced the analytical equation of the Bloch vector evolution of a two-level atoms subjected to a series of light pulses of different lengths, laser detunings, field amplitudes and phases. This allowed us to determine the error signal and shift of a probe-laser frequency locked on a narrow optical transition biased by a probe-induced shift using various interrogating schemes, as HR, MHR or GHR techniques. It is shown that these techniques do not allow a full cancelation of the shift in presence of dissipative processes. We have then proposed new universal protocols based on composite pulses and $\pm\pi/4$ and $\pm3\pi/4$  phase modulation.
The synthesized laser frequency locking-point is absolutely robust against pulse area errors and uncompensated probe-induced frequency-shifts in presence of laser induced decoherence and relaxation caused by both spontaneous emission and weak collisions. This is the first time that a composite laser-pulses interrogation protocol demonstrates a very efficient elimination of fields-induced frequency-shifts with non interacting particles through large constraints in laser parameters. These schemes can be implemented in two flavours: either by inverting clock state initialization or by pulse order reversal and are still competitive compared with HR and MHR schemes to a systematic imperfection in laser phase-stepping process during the error signal reconstruction.
We have indeed not considered other important technical problems such as local oscillator phase-noise, rapid laser power fluctuation
or electronic servo bandwidth restriction which are out of the scope of this paper. However such noise sources should reduce the locked frequency stability but not necessarily its accuracy.

Our frequency measurement protocol might be applied to weakly allowed or forbidden atomic transitions and might be very useful for the next generation of
1D and 3D optical lattice clocks \cite{Akatsuka:2010,Campbell:2017} probed by direct laser excitation or by high power magic-wave induced transitions \cite{Ovsiannikov:2007},
magnetically induced spectroscopy \cite{Barber:2006,Baillard:2007,Kulosa:2015} or based on Hyper-Raman Ramsey spectroscopy \cite{Zanon:2014}.
Laser spectroscopy protected against probe-field-induced frequency-shifts will perform better high-resolution frequency measurements by suppressing
spurious phase-shifts from the excitation pulses in precision spectroscopy \cite{Arnoult:2010,Matveev:2013}, Doppler-free two-photon spectroscopy \cite{Cohen-Tannoudji:1977,Salour:1977,Borde:1983}, tracking the tiniest changes in molecular vibrational frequencies based on clocks sensitive to potential variation in the electron-to-proton mass ratio \cite{Derevianko:2012,Safronova:2014,Schiller:2014,Yudin:2014}, fundamental physics tests and metrology with hydrogen molecular ions \cite{Karr:2016}, future nuclear clocks based on $\gamma$ transitions \cite{Campbell:2012,Peik:2015}, to observe some unexpected clock frequency shifts related to mass defect effects \cite{Yudin:2017} and in the recent application of Ramsey-type mass spectrometry \cite{Eibach:2011,George:2011}.

Thus, a new generation of optical generalized hyper-Ramsey quantum clocks may achieve a unprecedented breakthrough in extreme precision measurements for the next targeted $10^{-19}$ level of relative accuracy.

\section*{ACKNOWLEDGMENTS}

T. Zanon-Willette deeply acknowledges E. de Clercq, E. Arimondo, C.J. Bordé, B. Darquié, M. Glass-Maujean, C. Janssen, M.H. Levitt, A. Ludlow, Y. Té and J. Ye for suggestions and careful reading of the manuscript.

V.I.Yudin was supported by the Ministry of Education and Science of the Russian Federation (Project No. 3.1326.2017).
A.V. Taichenachev was supported by the Russian Scientific Foundation (Project No. 16-12-00052).

\subsection*{APPENDIX A: TIME-DEPENDENT MATRIX ELEMENTS}

\indent The analytic solution of generalized composite laser-pulses used to design
our universal protocol is explicitly expressed along with an in-depth analysis of the error signal construction and how the proposed protocol exploits
symmetrization to provide robustness against probe-induced frequency-shifts and various dissipative processes.
Some important results based on a combination of specific phase-modulated (GHR) resonances realize a very robust clock laser stabilization scheme against decoherence.
The generalized Hyper-Ramsey resonance is described within the density matrix formalism including decoherence.
The optical Bloch equations presented in the main text (see Eq.~(\ref{set-Bloch-DM})) describe the laser field interaction with a two-state quantum system. The general solution $\textup{M}(\theta_{l})$ is derived in a matrix form including $\textup{M}_{l}(\infty)$ steady-state solutions written as \cite{Jaynes:1955,Schoemaker:1978}:
\begin{equation}
\begin{split}
\textup{M}(\theta_{l})&=\textup{R}(\theta_{l})\left[\textup{M}_{l}(0)-\textup{M}_{l}(\infty)\right]+\textup{M}_{l}(\infty),\\
\textup{M}_{l}(\infty)&=-\frac{\Gamma}{\mathcal{D}}\left(\begin{array}{ccc}
                                                          \delta_{l}\Omega_{l}\cos\varphi_{l}-\gamma_{c}\Omega_{l}\sin\varphi_{l}\\
													     \gamma_{c}\Omega_{l}\cos\varphi_{l}+\delta_{l}\Omega_{l}\sin\varphi_{l} \\
													   \gamma_{c}^{2}+\delta_{l}^{2}
\end{array}\right),\\
\mathcal{D}&=\gamma_{c}\Omega_{l}^{2}+(\Gamma+2\xi)(\gamma_{c}^{2}+\delta_{l}^{2}).\\
\end{split}
\label{Jaynes-solution}
\end{equation}
where the clock frequency detuning is defined by $\delta_{l}=\delta-\Delta_{l}$ ($\Delta_{l}$ is the uncompensated part of the probe-induced frequency-shift) and the generalized pulse area is $\theta_{l}=\omega_{l}\tau_{l}$.
The square evolution matrix $\textup{R}(\theta_{l})$ requires exponentiation of the $\beta_{l}$ matrix
(Eq.~(3)). These square matrix elements $\textup{R}_{\textup{mn}}(\theta_{l})$ following refs \cite{Schoemaker:1978,Berman:2011} are given by:
\begin{equation}
\begin{split}
\textup{R}_{11}(\theta_{l})=e^{-\gamma_{c}\tau_{l}}&\left(a_{0}-a_{2}[\delta_{l}^{2}+\Omega_{l}^{2}\sin^{2}\varphi_{l}]\right),\\
\textup{R}_{12}(\theta_{l})=e^{-\gamma_{c}\tau_{l}}&\left(a_{1}\delta_{l}+a_{2}\Omega_{l}^{2}\sin\varphi_{l}\cos\varphi_{l}\right),\\
\textup{R}_{13}(\theta_{l})=e^{-\gamma_{c}\tau_{l}}&\left(a_{2}[\delta_{l}\Omega_{l}\cos\varphi_{l}-\Delta\gamma\Omega_{l}\sin\varphi_{l}]\right.\\
&\left.-a_{1}\Omega_{l}\sin\varphi_{l}\right),\\
\textup{R}_{21}(\theta_{l})=e^{-\gamma_{c}\tau_{l}}&\left(-a_{1}\delta_{l}+a_{2}\Omega_{l}^{2}\sin\varphi_{l}\cos\varphi_{l}\right),\\
\textup{R}_{22}(\theta_{l})=e^{-\gamma_{c}\tau_{l}}&\left(a_{0}-a_{2}[\delta_{l}^{2}+\Omega_{l}^{2}\cos^{2}\varphi_{l}]\right),\\
\textup{R}_{23}(\theta_{l})=e^{-\gamma_{c}\tau_{l}}&\left(a_{2}[\delta_{l}\Omega_{l}\sin\varphi_{l}+\Delta\gamma\Omega_{l}\cos\varphi_{l}]\right.\\
&\left.+a_{1}\Omega_{l}\cos\varphi_{l}\right),\\
\textup{R}_{31}(\theta_{l})=e^{-\gamma_{c}\tau_{l}}&\left(a_{2}[\delta_{l}\Omega_{l}\cos\varphi_{l}+\Delta\gamma\Omega_{l}\sin\varphi_{l}]\right.\\
&\left.+a_{1}\Omega_{l}\sin\varphi_{l}\right),\\
\textup{R}_{32}(\theta_{l})=e^{-\gamma_{c}\tau_{l}}&\left(a_{2}[\delta_{l}\Omega_{l}\sin\varphi_{l}-\Delta\gamma\Omega_{l}\cos\varphi_{l}]\right.\\
&\left.-a_{1}\Omega_{l}\cos\varphi_{l}\right),\\
\textup{R}_{33}(\theta_{l})=e^{-\gamma_{c}\tau_{l}}&\left(a_{0}+a_{1}\Delta\gamma-a_{2}[\Omega_{l}^{2}-\Delta\gamma^{2}]\right),
\end{split}
\label{Bloch-matrix-components}
\end{equation}
where $\Delta\gamma=\gamma_{c}-(\Gamma+2\xi)$.
Auxiliary time-dependent functions $\textup{a}_{0}\equiv\textup{a}_{0}(\theta_{l}),\textup{a}_{1}\equiv\textup{a}_{1}(\theta_{l}),\textup{a}_{2}\equiv\textup{a}_{2}(\theta_{l})$ are given by \cite{Torrey:1949,Schoemaker:1978}:
\begin{equation}
\begin{split}
a_{0}(\theta_{l})=&[(\textup{SD}_{3}-\textup{TD}_{2})\sin\theta_{l}+(\textup{SD}_{2}+\textup{TD}_{3})\cos\theta_{l}]e^{\rho_{l}\tau_{l}}\\
&+(\textup{D}_{0}\eta_{l}+\textup{g}_{l}^{2})\textup{R}e^{\eta_{l}\tau_{l}},\\
a_{1}(\theta_{l})=&[(\textup{SD}_{1}-\textup{T}\omega_{l})\sin\theta_{l}+(\textup{S}\omega_{l}+\textup{TD}_{1})\cos\theta_{l}]e^{\rho_{l}\tau_{l}}\\
&+\textup{D}_{0}\textup{R}e^{\eta_{l}\tau_{l}},\\
a_{2}(\theta_{l})=&[\textup{S}\sin\theta_{l}+\textup{T}\cos\theta_{l}]e^{\rho_{l}\tau_{l}}+\textup{R}e^{\eta_{l}\tau_{l}},
\end{split}
\label{auxiliary-functions}
\end{equation}
and relations between derivatives as \cite{Schoemaker:1978}:
\begin{equation}
\begin{split}
\dot{a}_{0}(\theta_{l})&=\delta_{l}^{2}\Delta\gamma a_{2}(\theta_{l}),\\
\dot{a}_{1}(\theta_{l})&=a_{0}(\theta_{l})-g_{l}^{2}a_{2}(\theta_{l}),\\
\dot{a}_{2}(\theta_{l})&=a_{1}(\theta_{l})+\Delta\gamma a_{2}(\theta_{l}),
\end{split}
\label{derivative-relation}
\end{equation}
with an auxiliary variable for convenience:
\begin{equation}
\begin{split}
a_{3}(\theta_{l})=a_{0}(\theta_{l})-a_{2}(\theta_{l})\delta_{l}^{2}.
\end{split}
\label{}
\end{equation}
We introduce the following notation:
\begin{equation}
\begin{split}
g_{l}^{2}=&\Omega_{l}^{2}+\delta_{l}^{2},\\
\textup{D}_{0}=&\eta_{l}-\Delta\gamma,\\
\textup{D}_{1}=&\rho_{l}-\Delta\gamma,\\
\textup{D}_{2}=&\omega_{l}(2\rho_{l}-\Delta\gamma),\\
\textup{D}_{3}=&(\rho_{l}^{2}-\omega_{l}^{2}-\rho_{l}\Delta\gamma+g_{l}^{2}),\\
\end{split}
\end{equation}
and
\begin{equation}
\begin{split}
\textup{R}=&\frac{1}{(\rho_{l}-\eta_{l})^{2}+\omega_{l}^{2}},\\
\textup{S}=&\frac{(\rho_{l}-\eta_{l})}{\omega_{l}((\rho_{l}-\eta_{l})^{2}+\omega_{l}^{2})},\\
\textup{T}=&\frac{-1}{(\rho_{l}-\eta_{l})^{2}+\omega_{l}^{2}}.
\end{split}
\label{integration-constants}
\end{equation}
The three roots of the matrix (one real root $\eta_{l}$ and two complex ones $\rho_{l}\pm i\omega_{l}$) are by Cardan's cubic solutions leading to damping terms $\eta_{l},\rho_{l}$ and a generalized angular frequency $\omega_{l}$ written as:
\begin{equation}
\begin{split}
\eta_{l}=&\frac{1}{3}(\Delta\gamma-\textup{C}-\frac{\Delta_{0}}{\textup{C}}),\\
\rho_{l}=&\frac{1}{3}(\Delta\gamma+\frac{\textup{C}}{2}+\frac{\Delta_{0}}{2\textup{C}}),\\
\omega_{l}=&\frac{\sqrt{3}}{6}(-\textup{C}+\frac{\Delta_{0}}{\textup{C}}),\\
\Delta_{0}=&\Delta\gamma^{2}-3g_{l}^{2},\\
\Delta_{1}=&-2\Delta\gamma^{3}+9g_{l}^{2}\Delta\gamma - 27\delta_{l}^{2}\Delta\gamma,\\
\textup{C}=&\sqrt[3]{\frac{\Delta_{1}+\sqrt{\Delta_{1}^{2}-4\Delta_{0}^{3}}}{2}}.
\end{split}
\label{cardan-roots}
\end{equation}

\subsection*{APPENDIX B: CLOCK FREQUENCY SHIFT $\delta\nu$ FROM RESONANCE LINE-SHAPES}

Using exact analytic expressions to solve the Bloch equations for a single given Rabi pulse, the
expression for a full sequence of $\textup{n}$ pulses, can be generalized to:
\begin{equation}
\begin{split}
\textup{M}(\theta_{1},...,\theta_{\textup{n}})=&\sum_{\textup{p}=1}^{\textup{n}}\left[\left(\overleftarrow{\prod_{l=\textup{p}}^{\textup{n}}}\textup{R}(\theta_{l})\right)\left(\textup{M}_{\textup{p}-1}(\infty)-\textup{M}_{\textup{p}}(\infty)\right)\right]\\
&+\textup{M}_{\textup{n}}(\infty).
\end{split}
\label{generalized-n-pulses}
\end{equation}
where state initialization means $\textup{M}_{0}(\infty)\equiv\textup{M}_{1}(0)$ by convention.
Such an expression can be rewritten to the canonical form presented in the paper (see Eq.~\ref{eq:canonical-form}) using some phasor expressions while fixing
index $\textup{k}$ for the free evolution time. Any composite laser-pulses sequence can indeed be re-casted such as:
\begin{equation}
\begin{split}
M(\theta_{1},...,\theta_{\textup{n}})\equiv\textup{A}+\textup{B}(\Phi)\cos(\delta\textup{T}+\Phi).
\end{split}
\label{eq:canonical-form-appendix}
\end{equation}
as long as we consider a unique pulse switching off the laser field used as a pivot in the factorization process.
The population transfer $P_{|\textup{g}\rangle\mapsto|\textup{e}\rangle}$ is related to the third component of the Bloch components $M(\theta_{1},...,\theta_{\textup{n}})$ as:
\begin{equation}
\begin{split}
P_{|\textup{g}\rangle\mapsto|\textup{e}\rangle}=\frac{1+\textup{W}(\theta_{1},...,\theta_{\textup{n}})}{2}.
\end{split}
\label{eq:population-transfer}
\end{equation}
To establish the frequency-shift of the resonance curve associated to the population transfer $P_{|\textup{g}\rangle\mapsto|\textup{e}\rangle}$, tracking the extremum of Eq.~\ref{eq:canonical-form-appendix} is required.
The condition is given by $\partial P_{|\textup{g}\rangle\mapsto|\textup{e}\rangle}/\partial\delta|_{\delta\rightarrow0}=0$ which leads to the first order expression as:
\begin{equation}
\begin{split}
\delta\nu=-\frac{\Phi|_{\delta\rightarrow0}}{2\pi\left(\textup{T}+\partial_{\delta}\Phi|_{\delta\rightarrow0}\right)},
\label{eq:High-order-clock-frequency-shift}
\end{split}
\end{equation}
where $\partial_{\delta}$ means a derivation with respect to the unperturbed clock detuning $\delta$. When high-order corrections are taken into account
in Eq.~\ref{eq:High-order-clock-frequency-shift}, the phase-shift has to be replaced by $\Phi\mapsto\Phi+\Psi+\Theta$ where:
\begin{subequations}
\begin{align}
\Psi&=-\arctan\left[\frac{\partial_{\delta}\textup{B}(\Phi)}{(\textup{T}+\partial_{\delta}\Phi)\textup{B}(\Phi)}\right]\label{eq:high-order-terms-a},\\
\Theta&=\arcsin\left[\frac{\partial_{\delta}\textup{A}}{\sqrt{\left[\partial_{\delta}\textup{B}(\Phi)\right]^{2}+\left[(\textup{T}+\partial_{\delta}\Phi)\textup{B}(\Phi)\right]^{2}}}\right].
\label{eq:high-order-terms-b}
\end{align}
\end{subequations}
High-order terms given by Eq.~\ref{eq:high-order-terms-a} and Eq.~\ref{eq:high-order-terms-b}
can handle a possible distortion of the line-shape when the free evolution time T is not so large compared to pulse duration.

\subsection*{APPENDIX C: CLOCK FREQUENCY SHIFT $\delta\widetilde{\nu}$ FROM ERROR SIGNAL LINE-SHAPES}

The error signal given by Eq.~\ref{eq:error-signal} used to lock the laser frequency is generated by taking the difference between two phase-modulated resonances as:
\begin{equation}
\begin{split}
\Delta\textup{E}=\textup{P}_{|\textup{g}\rangle\mapsto|\textup{e}\rangle}(\varphi_{l+})-\textup{P}_{|\textup{g}\rangle\mapsto|\textup{e}\rangle}(\varphi_{l-}).
\label{eq:error-signal-appendix}
\end{split}
\end{equation}
For instance the shift $\delta\nu$ of the frequency locking-point from the error signal due to an imperfect light-shift compensation is given by the relation:
\begin{equation}
\Delta\textup{E}|_{\delta=\delta\widetilde{\nu}}=0.
\label{eq:sensitivity-signal}
\end{equation}
To evaluate the clock frequency shift associated to different phase-step modulations, we use Eq.~\ref{eq:sensitivity-signal} to determine the analytical form of the frequency-shifted locking-point as:
\begin{equation}
\begin{split}
\delta\widetilde{\nu}=\frac{1}{2\pi \textup{T}}\left(-\widetilde{\Phi}|_{\delta\rightarrow0}\pm\arccos\left[-\frac{\widetilde{\textup{A}}|_{\delta\rightarrow0}}{\widetilde{\textup{B}}(\widetilde{\Phi})|_{\delta\rightarrow0}}\right] \right)
\end{split}
\label{eq:clock-frequency-shift-appendix}
\end{equation}
with a new phase-shift expression:
\begin{equation}
\begin{split}
\widetilde{\Phi}=\arctan\left[\frac{\textup{B}(\Phi)_{(\varphi_{l+})}\sin\Phi_{(\varphi_{l+})}-\textup{B}(\Phi)_{(\varphi_{l-})}\sin
\Phi_{(\varphi_{l-})}}{\textup{B}(\Phi)_{(\varphi_{l+})}\cos\Phi_{(\varphi_{l+})}-\textup{B}(\Phi)_{(\varphi_{l-})}\cos\Phi_{(\varphi_{l-})}}\right]
\end{split}
\label{eq:error-signal-phase-shift}
\end{equation}
including new offset and amplitude parameters as:
\begin{equation}
\begin{split}
\widetilde{\textup{A}}=&\textup{A}_{(\varphi_{l+})}-\textup{A}_{(\varphi_{l-})}\\
\widetilde{\textup{B}}(\widetilde{\Phi})=&\left[\textup{B}(\Phi)_{(\varphi_{l+})}\cos\Phi_{(\varphi_{l+})}-\textup{B}(\Phi)_{(\varphi_{l-})}\cos\Phi_{(\varphi_{l-})}
\right]\\
&\times\sqrt{1+\tan^2\widetilde{\Phi}}
\end{split}
\label{eq:error-signal-parameters}
\end{equation}

\subsection*{APPENDIX D: ERROR SIGNAL $\Delta\textup{E}_{\Downarrow\Uparrow}$($\Delta\textup{E}^{\dagger}_{\Downarrow\Uparrow}$) LINE-SHAPE}

\indent The proposed universal protocol interleaving $\pm\pi/4$ and $\pm 3\pi/4$ laser-phase steps
with a Bloch-vector initialization in each quantum state allows for an exact cancelation of all cosine terms in the
error signal pattern.
It is leaving a pure dispersive signal $\Delta\textup{E}\equiv\Delta\textup{E}[\textup{GHR}(\pi/4,3\pi/4)]$
while providing a perfectly robust locking-point at the unperturbed clock frequency $\delta=0$ against residual probe-induced frequency-shifts.
The error signal shape is evaluated explicitly based on a GHR protocol defined by 4 composite pulses $(\theta_{l})$ ($l=\textup{1,2,3,4}$)
where phase-steps are applied only within $\theta_{3}$ and with a free evolution time when $l=\textup{k}=2$ fixing $\theta_{2}=\delta\textup{T}$.
Due to pulse parameters that are defined by $\delta_{l}\equiv\delta-\Delta$ during laser interaction, $\Omega_{l}\equiv\Omega=\pi/2\tau$ and
choice of successive pulse durations as $\tau,T,2\tau,\tau$, a standard relation $R(\theta_{1})=R(\theta_{4})$ is obtained.

The dispersive error signal $\Delta\textup{E}$ based on Fig.~\ref{fig:GHR-protocol}(a) is then computed by successive differences between Bloch-vector
components alternating negative and positive $\varphi_{3}=\pi/4,3\pi/4$ phase-steps reducing to the compact expression:
\begin{equation}
\begin{split}
\Delta\textup{E}=&\textup{R}(\theta_{1})\Delta\textup{R}(\theta_{3})\textup{R}(\delta\textup{T})\textup{M}(\theta_{1}),
\end{split}
\label{eq:GHR-error-signal}
\end{equation}
where we have introduced:
\begin{equation}
\begin{split}
\Delta\textup{R}(\theta_{3})=&\textup{R}(\theta_{3})_{(+\frac{\pi}{4})}-\textup{R}(\theta_{3})_{(-\frac{\pi}{4})}\\
&-\left(\textup{R}(\theta_{3})_{(+3\frac{\pi}{4})}-\textup{R}(\theta_{3})_{(-3\frac{\pi}{4})}\right).
\end{split}
\label{eq:combination}
\end{equation}
Bloch-vector components initialization for the first pulse is here $\textup{M}_{1}(0)=(0,0,\textup{W}(0))$.
Note that successive differences between $\textup{M}_{l}$$_{(\pm\varphi_{l})}$ steady-states and from cross-product terms of the form
$\textup{R}(\theta_{3})$$_{(\pm\varphi_{l})}$$\textup{M}_{l}$$_{(\pm\varphi_{l})}$ cancel together exactly due to the particular choice of phase-steps.

When steady-states are vanishing $\textup{M}_{l}(\infty)\equiv0$, Eq.~(\ref{eq:GHR-error-signal}) can be directly reduced to the single product expression:
\begin{equation}
\begin{split}
\Delta\textup{E}=\textup{R}(\theta_{1})\Delta\textup{R}(\theta_{3})\textup{R}(\delta\textup{T})R(\theta_{1})\textup{M}_{1}(0).
\label{eq:GHR-error-signal-reduction}
\end{split}
\end{equation}
A symmetrization occurs for $\Delta\textup{R}(\theta_{3})$ and comes from exploiting $\pm\pi/4,\pm3\pi/4$ phase combinations between successive
sequences of composite laser-pulses, $\varphi \rightarrow-\varphi$ for cosine terms and $\varphi\rightarrow
\pi-\varphi$ for sine terms leading to a simple Pauli-like matrix:
\begin{equation}
\begin{split}
\Delta\textup{R}(\theta_{3})=2\textup{a}_{2}(\theta_{3})e^{-2\gamma_{c}\tau}\Omega^{2}\begin{array}{rcl}
	\begin{pmatrix}
0 & 1 & 0 \\
				1 & 0 & 0 \\
				 0 & 0 & 0
\end{pmatrix}
\end{array}
\end{split}
\label{eq:Pauli-matrix}
\end{equation}
The final compact expression of the error signal is rewritten as:
\begin{equation}
\Delta\textup{E}=\textup{A}\textup{W}(0)\left(
  \begin{array}{c}
 \textup{C}_{\textup{u}}\cos(\delta\textup{T})+\textup{S}_{\textup{u}}\sin(\delta\textup{T}) \\
 \textup{C}_{\textup{v}}\cos(\delta\textup{T})+\textup{S}_{\textup{v}}\sin(\delta\textup{T})  \\
 -\textup{S}_{\textup{w}}\sin(\delta\textup{T}) \\
  \end{array}
\right),
\label{eq:final-error-signal}
\end{equation}
where $\textup{A}=2a_{2}(\theta_{3})\Omega^{2}e^{-\gamma_{c}(4\tau+T)}$. Matrix elements $S_{\textup{u,v,w}}$ and $C_{\textup{u,v,w}}$ are reduced in a compact form using relations from Eq.~(\ref{derivative-relation}) as:
\begin{equation}
\begin{split}
\textup{S}_{\textup{u}}=&\Omega\delta_{1}\left[a_{1}(\theta_{1})\dot{a}_{2}(\theta_{1})-a_{3}(\theta_{1})a_{2}(\theta_{1})\right],\\
\textup{S}_{\textup{v}}=&\Omega\left[\dot{a}_{1}(\theta_{1})\dot{a}_{2}(\theta_{1})+a_{1}(\theta_{1})a_{2}(\theta_{1})\delta_{1}^{2}\right],\\
\textup{S}_{\textup{w}}=&\Omega^{2}\left[\left[\dot{a}_{2}(\theta_{1})\right]^{2}+a_{2}(\theta_{1})^{2}\delta_{1}^{2}\right],\\
\textup{C}_{\textup{u}}=&\Omega\left[a_{1}(\theta_{1})a_{2}(\theta_{1})\delta_{1}^{2}+a_{3}(\theta_{1})\dot{a}_{2}(\theta_{1})\right],\\
\textup{C}_{\textup{v}}=&\Omega\delta_{1}\left[\dot{a}_{1}(\theta_{1})a_{2}(\theta_{1})-a_{1}(\theta_{1})\dot{a}_{2}(\theta_{1})\right],\\
\textup{C}_{\textup{w}}=&0.
\end{split}
\label{eq:matrix-component-error-signal}
\end{equation}
with the unperturbed clock detuning corrected by probe-induced shifts as $\delta_{1}=\delta-\Delta$.

The normalized error signal connected to the third Bloch-vector component $\Delta\textup{E}_{\Downarrow(\Uparrow)}\equiv\Delta\textup{E(W)}_{\Downarrow(\Uparrow)}$ is extracted by taking differences between several population excitation fraction measurements. When only a decoherence term $\gamma_{c}$ is active, all steady-states are indeed vanishing.
The normalized error signal is given by Eq.~(\ref{eq:error-signal-decoherence}) with population initialization in either ground state $|g\rangle\equiv\Downarrow$ or excited state $|e\rangle\equiv\Uparrow$. We then have:
\begin{equation}
\begin{split}
\Delta\textup{E}_{\Downarrow(\Uparrow)}=-\frac{1}{4}\textup{A}_{\Downarrow(\Uparrow)}(0)\left[\dot{a}_{2}(\theta_{1})^{2}+a_{2}(\theta_{1})^{2}\delta_{1}^{2}\right]\sin(\delta\textup{T}).
\end{split}
\label{eq:error-signal-1}
\end{equation}
with $\textup{A}_{\Downarrow(\Uparrow)}(0)=\Omega^{2}\textup{A}\textup{W(0)}$ and where we apply $\textup{W(0)}_{\Downarrow(\Uparrow)}=-1$ ($+1$) respectively.
This is always a dispersive curve centered at the unperturbed optical clock frequency which is completely free from probe-induced frequency-shifts at all orders.

When decoherence term $\gamma_{c}$ and relaxation terms $\Gamma,\xi$ are simultaneously present, steady-states are non vanishing.
However, when two sets of Eq.~(\ref{eq:GHR-error-signal}) interleaved by population initialization in both states are applied,
the difference following Eq.~(\ref{eq:error-signal-relaxation-decoherence}) gives an identical error signal expression eliminating steady-states as:
\begin{equation}
\begin{split}
\Delta\textup{E}_{\Downarrow\Uparrow}=-\frac{1}{8}\textup{A}_{\Downarrow\Uparrow}(0)\left[\dot{a}_{2}(\theta_{1})^{2}+a_{2}(\theta_{1})^{2}\delta_{1}^{2}\right]\sin(\delta\textup{T}).
\end{split}
\label{eq:error-signal-2}
\end{equation}
with $\textup{A}_{\Downarrow(\Uparrow)}(0)=\Omega^{2}\textup{A}\left(\textup{W(0)}_{\Downarrow}-\textup{W(0)}_{\Uparrow}\right)$.
The resulting dispersive pattern versus the unperturbed clock frequency detuning $\delta$ is given by Eq.~(\ref{eq:error-signal-2}) taking $\textup{W(0)}_{\Downarrow(\Uparrow)}=-1(+1)$ for a full population inversion between quantum states.

It is also possible to read the sequence of composite pulses from left to right or from right to left by applying a
time reversal symmetry $t\rightarrow-t$ and phase inversion $\varphi\rightarrow-\varphi$ on diagram shown in
Fig.~\ref{fig:GHR-protocol}(a) leading to another equivalent scheme presented in Fig.~\ref{fig:GHR-protocol}(b).
We derive an alternative error signal called $\Delta\textup{E}^{\dagger}$, following the mirror-like protocol shown in Fig.~\ref{fig:GHR-protocol}(b).
We simply apply a permutation of laser parameters between pulse areas $\theta_{2}\leftrightarrow \theta_{3}$ stil keeping $R(\theta_{1})=R(\theta_{4})$
which directly leads to another error signal expression as:
\begin{equation}
\begin{split}
\Delta\textup{E}^{\dagger}=&-\textup{R}(\theta_{1})\textup{R}(\delta\textup{T})\Delta\textup{R}(\theta_{3})\textup{M}(\theta_{1}).
\end{split}
\label{eq:reverse-GHR-error-signal}
\end{equation}
We still generate the normalized error signal $\Delta\textup{E}^{\dagger}_{\Downarrow(\Uparrow)}$ following Eq.~(\ref{eq:error-signal-decoherence}) when only decoherence is present or the normalized error signal $\Delta\textup{E}^{\dagger}_{\Downarrow\Uparrow}$ following Eq.~(\ref{eq:error-signal-relaxation-decoherence}) when decoherence and relaxation are both activated.
We obtain error signal line-shapes that are identical to Eq.~(\ref{eq:final-error-signal}), Eq.~(\ref{eq:matrix-component-error-signal}),
Eq.~(\ref{eq:error-signal-1}) and Eq.~(\ref{eq:error-signal-2}).

\subsection*{APPENDIX E: ERROR SIGNAL $\Delta\textup{E}_{\Downarrow\Downarrow}$ LINE-SHAPE}

\indent A ultimate ultra-stable universal interrogation protocol interleaving $\pm\pi/4$ and $\pm 3\pi/4$ laser-phase steps
with a Bloch-vector initialization in only one single quantum state allows for an exact cancelation of all cosine terms in the
error signal pattern.
We show here that by combining two GHR protocols with sequence of composite pulses that are reversed in time ordering, as shown in Fig.~\ref{fig:GHR-protocol}(c),
similar dispersive shapes are recovered eliminating population initialization in the upper state.
We focus on the ultra stable error signal which relies on a combination of $\Delta\textup{E}$ and $\Delta\textup{E}^{\dagger}$
based on the protocol reported in Fig.~\ref{fig:GHR-protocol}(c).
We obtain a dispersive line-shape that does not require initialization population in both states,
even insensitive to non vanishing real and imaginary part of any initial optical coherence $\textup{U(0)},\textup{V(0)}\neq0$ when starting the interrogation protocol, as follows:
\begin{equation}
\begin{split}
\Delta\overline{\textup{E}}\equiv&\Delta\textup{E}+\Delta\textup{E}^{\dagger}=R(\theta_{1})\Delta\textup{R}(\theta_{3},\delta\textup{T})\textup{M}(\theta_{1}).
\end{split}
\label{eq:reverse-error-signal-relaxation-decoherence}
\end{equation}
where the commutator is $\Delta\textup{R}(\theta_{3},\delta\textup{T})=[\Delta\textup{R}(\theta_{3}),\textup{R}(\delta\textup{T})]$.
We derive exact expressions for matrix components as:
\begin{widetext}
\begin{equation}
\Delta\overline{\textup{E}}=A\left(
  \begin{array}{c}
  -\left[a_{1}(\theta_{1})^{2}\delta_{1}^{2}+a_{3}(\theta_{1})^{2}\right]\textup{U(0)}-\textup{S}_{1}\textup{V(0)}+\textup{S}_{2}\textup{W(0)}
   -\frac{\Gamma\Omega\delta_{1}}{\mathcal{D}}\left[-a_{3}(\theta_{1})\textup{S}_{4}+a_{1}(\theta_{1})\textup{S}_{5}\right] \\
   \textup{S}_{1}\textup{U(0)}+\left[a_{1}(\theta_{1})^{2}\delta_{1}^{2}+\dot{a}_{1}(\theta_{1})^{2}\right]\textup{V(0)}+\textup{S}_{3}\textup{W(0)}
   -\frac{\Gamma\Omega}{\mathcal{D}}\left[a_{1}(\theta_{1})\delta_{1}^{2}\textup{S}_{4}+\dot{a}_{1}(\theta_{1})\textup{S}_{5}\right] \\
   \textup{S}_{2}\textup{U(0)}-\textup{S}_{3}V(0)-\Omega^{2}\left[\dot{a}_{2}(\theta_{1})^{2}+a_{2}(\theta_{1})^{2}\delta_{1}^{2}\right]\textup{W(0)}
   +\frac{\Gamma\Omega^{2}}{\mathcal{D}}\left[a_{2}(\theta_{1})\delta_{1}^{2}\textup{S}_{4}+\dot{a}_{2}(\theta_{1})\textup{S}_{5}\right] \\
  \end{array}
\right)\sin(\delta\textup{T}).
\label{eq:combination-error-signal}
\end{equation}
\end{widetext}
with $A=4a_{2}(\theta_{3})\Omega^{2}e^{-\gamma_{c}(4\tau+T)}$.
We demonstrate that protocol shown in Fig.~\ref{fig:GHR-protocol}(c) is even more robust than protocols shown Fig.~\ref{fig:GHR-protocol}(a) and (b)
because all Bloch-vector matrix components are multiplied by a sine term eliminating uncompensated probe-induced frequency-shifts.\\
We introduce reduced variables $\textup{S}_{i}$ ($i=1,2,3,4,5$) as follows:
\begin{equation}
\begin{split}
\textup{S}_{1}=&a_{1}(\theta_{1})\delta_{1}\left[a_{3}(\theta_{1})-\dot{a}_{1}(\theta_{1})\right],\\
\textup{S}_{2}=&\Omega\delta_{1}\left[a_{1}(\theta_{1})\dot{a}_{2}(\theta_{1})-a_{2}(\theta_{1})a_{3}(\theta_{1})\right],\\
\textup{S}_{3}=&\Omega\left[a_{1}(\theta_{1})a_{2}(\theta_{1})\delta_{1}^{2}+\dot{a}_{1}(\theta_{1})\dot{a}_{2}(\theta_{1})\right],\\
\textup{S}_{4}=&e^{\gamma_{c}\tau}-a_{3}(\theta_{1})-a_{1}(\theta_{1})\gamma_{c}-a_{2}(\theta_{1})\left(\gamma_{c}^{2}+\delta_{1}^{2}\right),\\
\textup{S}_{5}=&a_{1}(\theta_{1})\delta_{1}^{2}+\gamma_{c}\left(e^{\gamma_{c}\tau}-\dot{a}_{1}(\theta_{1})\right)-\dot{a}_{2}(\theta_{1})\left(\gamma_{c}^{2}+\delta_{1}^{2}\right).
\end{split}
\label{eq:reduced-variables-matrix}
\end{equation}
We finally derive a new ultra stable normalized error signal $\Delta\overline{\textup{E}}_{\Downarrow\Downarrow}$
based on population transfer following Eq.~(\ref{eq:error-signal-relaxation-decoherence-reversion}) with $\textup{U(0)}=\textup{V(0)}=0$ as:
\begin{widetext}
\begin{equation}
\begin{split}
\Delta\overline{\textup{E}}_{\Downarrow\Downarrow}=-\frac{1}{8}\textup{A}\Omega^{2}\left[\left[\dot{a}_{2}(\theta_{1})^{2}+a_{2}(\theta_{1})^{2}\delta_{1}^{2}\right]
\textup{W(0)}_{\Downarrow\Downarrow}-\frac{\Gamma}{\mathcal{D}}\left[a_{2}(\theta_{1})\delta_{1}^{2}\textup{S}_{4}+\dot{a}_{2}(\theta_{1})\textup{S}_{5}\right]\right]\sin(\delta\textup{T}).
\end{split}
\label{eq:error-signal-3}
\end{equation}
\end{widetext}
The resulting dispersive pattern versus the unperturbed clock frequency
detuning $\delta$ is given by Eq.~(\ref{eq:error-signal-3}) taking only $\textup{W(0)}_{\Downarrow\Downarrow}=-1$.
If we neglect a small correction on signal contrast due to decoherence and relaxation terms in Eq.~(\ref{eq:error-signal-3}), we retrieve a line-shape expression
which is identical to Eq.~(\ref{eq:error-signal-1}) and Eq.~(\ref{eq:error-signal-2}) and does not require a population inversion between quantum states.

\end{document}